\documentclass{article}
\usepackage[utf8]{inputenc}
\usepackage[letterpaper,margin=1in]{geometry}

\usepackage{style_defaults}
\usepackage{macros}
\usepackage{authblk}
\usepackage{url}
\usepackage{hyperref}

\usepackage{booktabs, tabularx}
\usepackage{algorithmic, algorithm}

\title{Benchmarking Fraud Detectors on Private Graph Data}
\author{Alexander Goldberg}
\author{Giulia Fanti}
\author{Nihar B. Shah}
\author{Zhiwei Steven Wu}
\affil{Carnegie Mellon University}
\affil{\texttt {\{akgoldbe, gfanti, nihars, zhiweiw\}@andrew.cmu.edu}}
\date{}

\begin{document}

\maketitle

\begin{abstract}
We introduce the novel problem of benchmarking fraud detectors on private graph-structured data. Currently, many types of fraud are managed in part by automated detection algorithms that operate over graphs. We consider the scenario where a data holder wishes to outsource development of fraud detectors to third parties (e.g., vendors or researchers). The third parties submit their fraud detectors to the data holder, who evaluates these algorithms on a private dataset and then publicly communicates the results. We propose a realistic privacy attack on this system that allows an adversary to de-anonymize individuals' data based only on the evaluation results. In simulations of a privacy-sensitive benchmark for facial recognition algorithms by the National Institute of Standards and Technology (NIST), our attack achieves near perfect accuracy in identifying whether individuals’ data is present in a private dataset, with a True Positive Rate of 0.98 at a False Positive Rate of 0.00. We then study how to benchmark algorithms while satisfying a formal \emph{differential privacy (DP)} guarantee. We empirically evaluate two classes of solutions: subsample-and-aggregate and DP synthetic graph data. We demonstrate through extensive experiments that current approaches do not provide utility when guaranteeing DP. Our results indicate that the error arising from DP trades off between bias from distorting graph structure and variance from adding random noise. Current methods lie on different points along this bias-variance trade-off, but more complex methods tend to require high-variance noise addition, undermining utility. 
\end{abstract}

\section{Introduction}

\begin{figure*}[ht]
    \centering
    \begin{subfigure}[b]{0.48\linewidth}
        \centering
        \includegraphics[width=\linewidth, height=5cm]{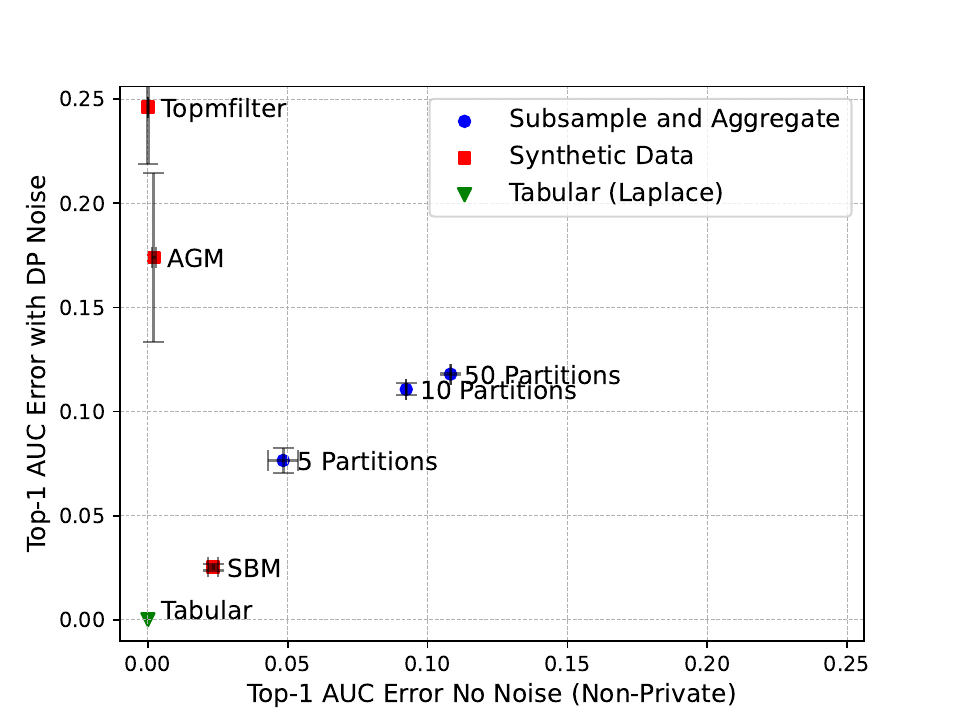}
        \caption{Amazon}
    \end{subfigure}%
    ~ 
    \begin{subfigure}[b]{0.48\linewidth}
        \centering
        \includegraphics[width=\linewidth, height=5cm]{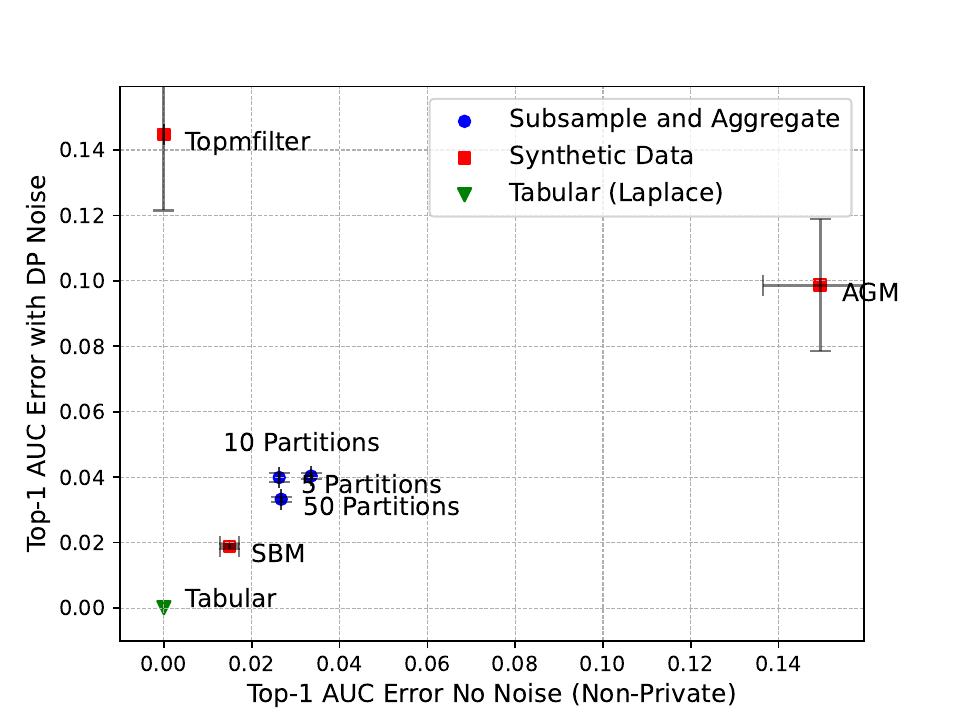}
        \caption{Yelp}
    \end{subfigure}
    \caption{Comparison of DP benchmarking methods for releasing the best AUC score among 10 fraud detectors with privacy budget of $\eps=5.0$. The horizontal axis captures error due to inductive bias (i.e., the underlying graph model, without DP noise); the vertical axis captures error including DP noise. More complex synthetic data methods (Topmfilter and AGM) can model the data without privacy, but suffer from high variance due to DP noise addition, undermining utility. Subsample-and-aggregate distorts graph structure extensively, even before adding random noise to outputs. All current methods incur large utility cost on graph data compared to tabular data. Error bars show the standard error of the mean across $10$ simulations of each method.}
   \label{fig:head_to_head_primary}
\end{figure*}

Fraud constitutes a pernicious problem across numerous domains, manifesting as fake product reviews, fraudulent payments, and the resale of stolen goods, among other harms \cite{ftc-fraud}. The scale of fraud losses is driven in part by the difficulty of detecting fraud: today, the problem is primarily handled by automated detectors with high false positive and false negative rates. Although many organizations dedicate entire teams to fraud detection, other organizations outsource the development of fraud detection mechanisms to third parties, such as vendors of fraud detection software and/or third-party researchers \cite{datavisor,riskified,kount}. However, effective outsourcing requires enterprises to share internal fraud data, which can be challenging due to privacy regulations (e.g., GDPR) and/or the risk of leaking trade secrets through shared datasets. As a result, the lack of publicly shareable data has limited research progress on detection of fraudulent behaviors in privacy-sensitive domains. For example, in scientific peer review, there is a lack of real data on reviewer-paper assignments. This limits researchers' ability to evaluate solutions to the problem of detecting rings of colluding reviewers~\cite{littman2021collusion, jecmen2020mitigating},  necessitating reliance on laboratory-generated~\cite{jecmen2023dataset} or semi-synthetic~\cite{jecmen2024detection} data.

\paragraph{Problem Statement.}

In this work, we explore a paradigm for outsourcing fraud detection in which sensitive data remains within an organization’s boundaries. Rather than sharing data externally, the organization invites third parties to submit fraud detection algorithms—leveraging domain knowledge and public data—which are then evaluated and ranked by the data holder. These third parties may be incentivized through financial or reputational rewards, such as leaderboard-based competitions~\cite{kaggle}. This paradigm has gained traction as a practical solution for enabling public evaluation on privacy-sensitive data. In recent years, it has been adopted in several high-profile applications, including fraud detection for Amsterdam’s social welfare system~\cite{amsterdam2025}, biometric identification benchmarking in India~\cite{uidai2025}, and facial recognition evaluation by the U.S. National Institute of Standards and Technology (NIST)~\cite{nist2025}. We study this setting under two key constraints that have not been jointly explored before:
\begin{enumerate}[leftmargin=*]
    \item \emph{Private algorithm evaluation}: We observe that if the accuracy of a fraud detector is released directly, it can leak sensitive information about the underlying test data (Section~\ref{sec:privacy-risk}). We therefore consider methods for evaluating algorithms, and releasing their results, under a differential privacy (DP) constraint \cite{dwork2006_dp}. 
    \item \emph{Graph-structured data}: Many prominent fraud domains, such as financial fraud or product review fraud, have graph-structured datasets. We focus on fraud detection algorithms (and privacy solutions) that can be applied to graph-structured data.
\end{enumerate}

More precisely, we consider a \emph{benchmarking server}, which has a private graph $\graph$ consisting of a set of known fraudulent vertices $\fraudNodes$ (of size $\nfraud$) and a set of benign vertices $\benignNodes$ (of size $\nbenign$). 

The benchmarking server's goal is to \emph{evaluate} one or more fraud detection algorithms and \emph{communicate} the result back to the algorithm designers. The benchmarking server receives a fraud detection algorithm $\fraudAlgo$. The fraud detection algorithm takes as input a vertex $v$ and the entire graph $\graph$ and outputs $\fraudAlgo(\graph, v)$ which is a numerical score where a higher score indicates a higher likelihood of fraud. For example, the fraud detection algorithm could score a vertex by its degree.
The benchmarking server returns an \emph{accuracy statistic} for the fraud detection algorithm on graph $\graph$. Concretely, we consider the \emph{AUC score}, which is defined as:
\[\auc(\fraudAlgo, \graph) =  \frac{1}{\nfraud \nbenign} \sum_{v_0 \in \benignNodes} \sum_{v_1 \in \fraudNodes} \ind[\fraudAlgo(\graph, v_1) > \fraudAlgo(\graph, v_0)]. \] The AUC score represents the probability that a randomly chosen fraudulent vertex is scored higher than a randomly chosen benign vertex. It is a commonly used accuracy statistic for class-imbalanced binary classification problems like fraud detection \cite{grover2022_fdb}.

\paragraph{Challenges and Approach.}
Existing techniques for differentially private release of statistics cannot be easily applied to graph-structured data (Section \ref{sec:graph_challenges}). 
The main challenge is that benchmarking fraud detection algorithms on graph-structured data relies on high-sensitivity queries over the graph, meaning that the query result can change significantly if even a single node's neighbors are altered in the graph (Definition \ref{defn:sensitivity}). Making such algorithms DP requires large amounts of noise, undermining utility.

The goal of this work is to instantiate and benchmark different classes of techniques for evaluating fraud detection algorithms over graph-structured data under a DP constraint. We evaluate two approaches for dealing with high-sensitivity queries: \emph{(1) Subsample-and-aggregate} partitions the dataset into non-overlapping datasets, then evaluates the fraud detectors over each partition. The average accuracy over the partitions is low-sensitivity, and can be released with less noise than without partitioning. \emph{(2) Synthetic graph data} generates a DP copy of the true graph; then, fraud detectors are evaluated on this synthetic graph. 

\paragraph{Contributions.}
Our primary contributions are:
\begin{enumerate}[leftmargin=*]
    \item We formulate the problem of \emph{differentially private benchmarking of fraud detectors on private graph data.} \textbf{We describe a simple privacy attack on a system that benchmarks user-submitted algorithms on private data.} Our attack applies broadly to benchmarking frameworks that evaluate algorithms on the entire dataset at once rather than per-user or instance-level evaluation---such as graph data or authentication systems where access to the full dataset is required for evaluation. \textbf{In simulations of a deployed facial recognition benchmarking system, we show that this attack is practical}---concretely, our  attack achieves near perfect accuracy in identifying whether individuals' data is present in a private dataset, with a True Positive Rate of 0.98 at a False Positive Rate of 0.00.
    \item  We then evaluate the potential of differential privacy as a solution concept for preserving privacy of graph data used in a benchmarking system. Across methods, \textbf{we observe a severe trade-off between bias introduced by distorting the graph and noise required to compensate for computing high sensitivity statistics on the graph.} This result is captured in Figure~\ref{fig:head_to_head_primary}, which shows the error in privately benchmarking the best AUC score among a set of $10$ fraud detectors. We plot the error of each DP benchmarking method without noise added (inductive bias) against error after adding noise to ensure differential privacy. Among synthetic data methods, more complex methods (TopmFilter and AGM) have lower inductive bias, but much higher noise addition to preserve privacy than the simpler SBM. Subsample-and-aggregate tends to distort graph structure extensively, even before adding random noise to outputs, but then has low additional error from the random noise. All current methods to satisfy DP on graph data incur large utility cost compared to tabular data.  
    \item To explain these results, we conduct detailed ablations on both subsample-and aggregate and synthetic data methods. While these methods introduce inductive bias in different ways, they exhibit a similar trade-off --- the less we bias our graph representation, the more noise we must add to satisfy DP.
\end{enumerate}

Our code is available at \href{https://github.com/akgoldberg/private_fraud_benchmarking}{\texttt{https://github.com/akgoldberg/private\_fraud\_benchmarking}}.

\section{Privacy Risk}
\label{sec:privacy-risk}

We start by describing an attack that a malicious actor can use to compromise the privacy of an individual included in the dataset used for algorithmic benchmarking. 

\subsection{Privacy Attack}

 Consider a bad actor who wishes to answer a binary query, such as whether an edge exists between two vertices in the graph. The adversary needs three capabilities. (1) \emph{An accurate fraud detector:} for example, a known algorithm from the literature which does better than random (AUC $> 0.5$). (2) \emph{An inaccurate fraud detector:} for example, scoring vertices at random (expected AUC $=0.5$). (3) \emph{The ability to identify vertices in $\graph$:} this depends on what information the private server gives to the fraud detection algorithm.\footnote{For a general binary query, the attacker would need an accurate estimator for that query given the private graph dataset.} In many cases, $\graph$ may include extensive metadata per vertex, which makes it easy to identify vertices. Even without metadata, there are many de-anonymization attacks leveraging only the graph structure (see \cite{shouling_graphdeanonsurvey} for a survey), which enable an adversary to identify vertices. 

\begin{algorithm}
\caption{Attacker's Submission to the Benchmarking Server}
\label{alg:privacy_attack}
\begin{algorithmic}[1]
\REQUIRE Accurate fraud detector $\fraudAlgo$, pair of vertices $v_1, v_2$.
\STATE Check if an edge exists between $v_1$ and $v_2$ in private graph $\graph$.
\IF{edge $(v_1, v_2)$ exists}
    \STATE Run accurate fraud detection algorithm $\fraudAlgo$ on $\graph$.
\ELSE
    \STATE Return a random fraud label for each vertex in $\graph$.
\ENDIF
\end{algorithmic}
\end{algorithm}

The adversary submits a ``fraud detector'' to the benchmark described in Algorithm~\ref{alg:privacy_attack}. The adversary identifies the relevant pair of vertices, and runs the accurate fraud detector if an edge exists between the vertices or an inaccurate fraud detector otherwise. If the adversary observes a high AUC score, they learn that an edge exists, while if they observe a low AUC score they learn that the edge does not exist. In effect, the benchmark allows a malicious actor to answer any binary query on $\graph$ by encoding the query as either a high-accuracy or low-accuracy fraud detector. The privacy attack we describe above applies to a range of benchmarking systems that evaluate user-submitted algorithms on private data, as demonstrated in the next section.

\subsection{Attacking a Real-World Benchmark}
 To demonstrate the practical implications of the attack, we take an existing privacy-sensitive application as a case study: the National Institute of Standards and Technology (NIST) Face Recognition Technology Evaluation (FRTE). The FRTE benchmarks algorithms for facial recognition on sensitive private datasets of face images like mugshots, visa applicants, and border-crossing photos. The FRTE benchmarks the task of ``1:N face identification''. A 1:N face identification algorithm matches a given ``probe image'' against a large ``gallery dataset'' of images, returning an image in the gallery of the same person. While the dataset consists of face images, not a graph, the same attack proposed for graph data can be applied to the face recognition benchmark system. 

Specifically, an attacker with access to a reasonably accurate facial recognition algorithm can exploit the benchmarking system to determine whether one or more specific individuals' faces are included in the gallery dataset; for example, suppose that an attacker wants to know if Bob is in the gallery dataset. The attacker obtains an image of Bob. When the attacker's submission searches the gallery dataset for a given probe image, the attacker first uses the accurate facial recognition algorithm to check if Bob's image matches any image in the gallery dataset. If yes, they use the accurate algorithm on the \emph{actual probe image} (i.e., not Bob). If Bob is not in the gallery, they use the inaccurate algorithm on the probe image. A high accuracy score on the benchmark implies Bob’s presence, while a low score indicates his absence.\footnote{
We have disclosed this vulnerability to NIST, which has implemented steps to reduce its exploitability. Moreover, note that this attack only reveals membership in the dataset---it does not reveal other information about the individuals in the gallery, such as date of the photo or biographic information.
}

\subsection{Effectiveness of the Attack}

\begin{figure}[t]
    \centering
    \includegraphics[width=0.6\linewidth]{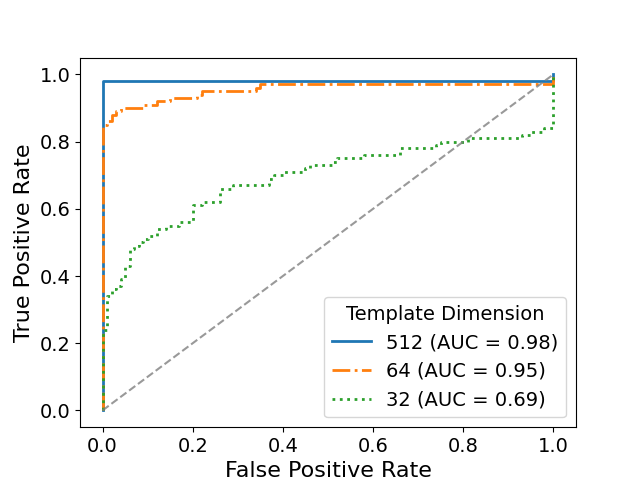}
    \caption{ROC curve of the privacy attack on NIST's FRTE benchmark, simulated on the CelebA dataset.}
    \label{fig:attack_roc_curve}
\end{figure}

To evaluate the practical viability of the attack, we simulate the 1:N face recognition benchmark using the publicly available CelebA dataset, which contains faces of over 10,0000 celebrities~\cite{celeba}. We use an open-source, deep learning-based facial recognition model ArcFace~\cite{deng2019arcface} as the accurate  model.
The attacker uses the face recognition model to generate embeddings (templates) of images and then performs ``identification'' using cosine similarity between embeddings. We vary the adversary's capabilities by reducing the dimension of the embeddings used by the model from 512 to 64 and 32. Then, true positive and false positive rates of the attack can be computed by varying the threshold at which the attack concludes that an attack image is present in the private data. In Figure~\ref{fig:attack_roc_curve}, we show the ROC curve of the attack. \textbf{Using 512-dimensional embeddings, the attack achieves a TPR of $0.98$ at an FPR of $0.00$, successfully identifying 98 out of 100 gallery members while avoiding false matches.} Even with 64- or 32-dimensional embeddings, the attack remains effective, achieving high AUC scores. This result highlights that the attack is feasible even on the FRTE benchmark using simple, open-source facial recognition models.\footnote{Our code and a detailed FRTE benchmark description are available at {\href{https://github.com/akgoldberg/face_recognition_privacy_attack}{\texttt{https://github.com/akgoldberg/face\_recognition\_privacy\_attack}}}.}

The FRTE privacy vulnerability demonstrates how publicly releasing algorithmic benchmarking results can compromise the anonymity of individuals in private datasets. In the remainder of this paper, we focus on benchmarks where the private data used for evaluation is a graph, as graph datasets are common in fraud detection tasks and pose significant technical challenges.

\section{Problem Formulation}

Based on the privacy risk posed by the attack in Section \ref{sec:privacy-risk}, we ask how to protect dataset privacy for the benchmarking server. We consider three different operating modes for the benchmarking server: (1) \emph{one-shot}: the server releases the AUC score for a single submitted fraud detector, (2) \emph{full leaderboard:} the server returns the AUC score for a set of submitted fraud detectors, and (3) \emph{top-1 release:} the server releases the best-performing fraud detector among a set of submitted algorithms.

\subsection{Incorporating Differential Privacy}
\label{sec:privacy_defn}

We propose that the fraud benchmarking server satisfy a relaxation of differential privacy (DP) \cite{dwork2006_dp} that protects benign vertices: 
\begin{definition}[Protected differential privacy \cite{kearns_roth_wu_protected_search}]
\label{defn:protected_dp}
Two graphs $\graph$, $\graph'$ are neighboring if:
\begin{enumerate}[label=(\alph*)]
    \item $\graph$ and $\graph'$ share the same partitions of fraudulent and non-fraudulent vertices $\fraudNodes$ and $\benignNodes$.
    \item $\graph$ can be obtained from $\graph'$ by rewiring the edges of one \emph{benign} vertex and/or changing that vertices' metadata.
\end{enumerate}
Let $\server$ denote the benchmarking server that given a graph and fraud detector outputs an estimate of the AUC score. The server $\server$ satisfies $\eps$-protected differential privacy if for any two neighboring graphs $\graph, \graph'$, any fraud detection algorithm $\fraudAlgo$ and any possible set of outputs $\mathcal{O}$:
\[\Pr[f(\graph, \fraudAlgo) \in \mathcal{O}] \leq e^{\eps} \Pr[f(\graph', \fraudAlgo) \in \mathcal{O}]. \]
\end{definition}

The definition of protected DP is identical to the standard definition of differential privacy, except in how ``neighboring'' graphs are defined. Specifically, 
Standard DP allows $\graph$ and $\graph'$ to differ in the data of \emph{any} vertex in the graph, not just a benign vertex. Protected DP is a relaxation of standard DP in that any graphs that are neighbors per the definition of protected DP are also neighbors under standard DP. 
Any mechanism that satisfies standard DP also satisfies protected DP. We 
will drop ``protected'' and refer to protected differential privacy as DP for brevity throughout.

We primarily adopt this relaxed notion of privacy to improve utility. In many real-world graphs, the rate of fraud is low. Hence, requiring that the released accuracy statistic does not change much if we change the connections of these fraudulent  vertices makes it difficult to release high-fidelity benchmarks. Still, we believe this relaxation is useful. In fraud detection, it is natural to hold different privacy expectations for fraudulent participants (many of which may even be fake \cite{jecmen2024_detection}) compared to legitimate ones. 
For example, in online review platforms or social networks, fraudsters can correspond to non-human bots who do not have the same privacy status as humans. 
Second, in domains where ground truth for fraud does not currently exist, a common approach is to inject simulated fraudulent vertices into a real graph of benign vertices \cite{jecmen2024_detection}. 
As these fraudulent vertices are simulated, they do not require privacy protection. 
Finally, we note that the benchmarking server can still provide baseline protections for fraudulent vertices, like not exposing personally identifiable information (PII), while guaranteeing stronger DP protection for benign vertices.

In the definition of neighboring graphs, we adopt the strong notion of \emph{node} differential privacy, which protects all of the edges of any single benign vertex. Many prior works employ a weaker notion of \emph{edge} differential privacy \cite{karwa_edgedp, chen2019_asg, jorgensen2016_asg, nguyen2015_topm}, which defines neighboring graphs as graphs that differ in a single edge. 
Node-DP gives a much stronger guarantee. In particular, the unit of participation in the dataset is at the vertex-level (i.e., whether an Amazon user's reviews are included in the dataset or not) and node-DP promises that the inclusion of a given individual in the dataset does not reveal much information about them. 
We note that protected DP inherits the \emph{composition property} of standard DP:
\begin{theorem}[Composition \cite{dwork2006_dp}]
    For any two fraud detectors $\fraudAlgo_1$ and $\fraudAlgo_2$, if releasing $\accuracy(\fraudAlgo_1, \graph)$ satisfies $\eps_1$-protected DP and releasing $\accuracy(\fraudAlgo_2, \graph)$ satisfies $\eps_2$-protected DP, then releasing both results on graph $\graph$, $( \accuracy(\fraudAlgo_1, \graph), \accuracy(\fraudAlgo_2, \graph))$ satisfies $(\eps_1 + \eps_2)$-DP.
    \label{thm:composition}
\end{theorem} 
This property is helpful in moving from one-shot release of fraud detectors to releasing a leaderboard of many fraud detectors.

\subsection{Challenges of Graph Data}
\label{sec:graph_challenges}

Even evaluating a single fraud detector on graph data proves challenging under DP constraints. To understand why, let us compare our setting to evaluating a fraud detector on tabular data. Evaluating a single fraud detector can be seen as a problem of releasing a (noisy) query result. A simple mechanism that solves the query release problem adds random noise with variance scaled to the ``sensitivity'' of this query, which is defined as follows.
\begin{definition}[Global Sensitivity]
    \label{defn:sensitivity}
    For a query $\accuracy: \mathcal{X} \to \R^d$, define its \emph{global sensitivity} \[\sensitivity_\accuracy = \max_{\graph, \graph' \text{ neighbors}} \| \accuracy(\graph) - \accuracy(\graph') \|_1 \] as the worst-case change in $\accuracy$ across any two neighboring graphs.
\end{definition}

Then, a canonical mechanism, termed the Laplace Mechanism, scales noise to the global sensitivity:

\begin{definition}[Laplace Mechanism \cite{dwork2006_dp}]
\label{defn:laplace}
     On any input $\graph$ the \emph{Laplace Mechanism} with privacy parameter $\eps$ releases \[ \widetilde{f}(\graph) = \accuracy(\graph) + \text{Laplace}(\sensitivity_\accuracy / \eps). \]
    The Laplace Mechanism satisfies $\eps$-DP.
\end{definition}

In the tabular setting, model evaluation is a low sensitivity query and therefore can be released by directly applying the Laplace mechanism. Consider a simple case where fraud detector $\fraudAlgo$ is a fitted logistic regression model (the weights of the model are fixed). Changing any row of a tabular dataset only changes the features of that row and hence changes at most a single fraud prediction score. Therefore, when evaluating the fixed model on tabular data, the AUC score can only change by $\frac{1}{\nbenign}$. The Laplace mechanism can then release the true AUC score of the fraud detector plus Laplace noise with variance $\frac{2}{(\eps \nbenign)^2}$.

In contrast, consider evaluating the logistic regression model on a graph where features of each vertex include graph statistics like the degree of each vertex. Because features of each vertex depend on other vertices, changing any one vertex can change features of all other vertices in the graph. In the worst-case, changing a vertex changes fraud prediction scores for all other vertices in the graph, so the AUC score has a large global sensitivity of $1$. As this is the largest possible value AUC can take, the Laplace mechanism must add so much noise that the entire signal is lost.

In this paper, we focus on addressing this challenge of high sensitivity of model evaluation on graph data. In cases where queries of a dataset have large worst-case sensitivity there are three classes of solutions in the DP literature:
\begin{enumerate}[leftmargin=*]
    \item (\emph{Subsample-and-aggregate}) Force low sensitivity of the AUC score by applying  ``subsample-and-aggregate.'' Partition the vertices of the graph into $\npartition$ equally sized subsets, compute AUC score on each partition and then, directly release the average AUC score across partitions plus noise.
    \item (\emph{Synthetic data}) Generate DP synthetic data that captures some structure of the private graph and run fraud benchmarking on this private graph data. 
    \item (\emph{Calibrate noise to ``local sensitivity.''}) Estimate (an upper bound) on how sensitive $\auc$ is on the specific graph and fraud detection algorithm $\fraudAlgo$ and calibrate noise to this sensitivity, which may be much lower than the worst case global sensitivity. This approach includes mechanisms like Propose-Test-Release, Smooth Sensitivity, and the Inverse Sensitivity Mechanism \cite{nissim_smooth_sensitvity_07, lei_2009} as well as recent work on privatizing black-box scripts run on private data \cite{kohli2023_blackboxdp}.
\end{enumerate}

In this work, we give instantiations of subsample-and-aggregate and synthetic data generation algorithms tailored to the benchmarking server setting and run extensive empirical evaluations to understand opportunities and shortcomings. We do not evaluate local sensitivity based methods, because these approaches are computationally infeasible in our setting as they would require enumerating every possible neighboring graphs and evaluating fraud detectors on these graphs to estimate a bound on local sensitivity. 
The rest of the paper is organized as follows. In Section~\ref{sec:related_work}, we survey related work on private model evaluation and on running arbitrary code on private data under DP constraints. In Section~\ref{sec:subsamp}, we describe our instantiation of the Subsample-and-Aggregate framework. In Section~\ref{alg:synthetic}, we survey existing methods for synthetic graph data and our choice of algorithms to benchmark. Finally, in Section~\ref{sec:methods} we detail the methodology of our extensive experiments and in Section~\ref{sec:results}, we analyze the results of these experiments.

\section{Related Work}
\label{sec:related_work}

To our knowledge, this work is the first to consider the problem of model evaluation on graph data under differential privacy constraints. For tabular (non-graph) data, there are two lines of work that consider DP model evaluation. One line of work (starting with \cite{reiter2009_verification}) proposes a framework of ``verification servers'' wherein analysts fit a model of data (e.g., a linear regression model) on a synthetic dataset and then employ a ``verification server'' which holds non-synthetic data to perform quality checks that their model is useful like goodness-of-fit tests. In subsequent work, they show how to design such a verification server for the special case of assessing quality of fit of a linear regression model \cite{yu2018_verification} and build a prototype for data on US government employees \cite{barrientos2021_validationserver}. This work validates the usefulness of a ``synthetic data plus verification server'' model for allowing public use of private datasets. However, these works focus on tabular data rather than graph data, which poses specific challenges as we detail in Section~\ref{sec:graph_challenges}.

A recent line of work in DP machine learning (starting with \cite{liu2018_privateselection} and extended in \cite{cohen2022_selection, papernot2021_hyperparameter}), looks at a closely related problem of model selection under differential privacy constraints. This work assumes the ability to train and evaluate a single model with $\eps$-DP, which is reasonable for the tabular-data setting where algorithms like DP-SGD are effective at model training and the Laplace mechanism (or Gaussian mechanism) solves-model evaluation. These works focus on choosing the (nearly) optimal model in minimizing loss among a large set of models without paying for privacy loss that grows with the number of models. In our work, we observe that on graphs (even ignoring model training) the seemingly straightforward step of one-shot model evaluation is difficult under differential privacy constraints. The methods for private selection could be applied in conjunction with our proposals for one-shot DP release.

A number of works have considered the problem of running arbitrary queries on private data. Subsample-and-aggregate, first proposed in \cite{nissim_smooth_sensitvity_07}, is one popular method for reducing the sensitivity of a query.
A practical instance of this framework was implemented in GUPT \cite{mohan2012_gupt}, which allows researchers to run arbitrary scripts on private data using subsample-and-aggregate. The paper shows that their system can enable researchers to fit models like $k$-means clustering and logistic regression on a (tabular) dataset of chemical compounds and achieve utility close to non-private for privacy budgets as small as $\eps = 2$. More recently, subsample-and-aggregate has been applied in the context of training machine learning models \cite{papernot2018_pate} via the ``PATE'' framework. PATE trains a ``teacher'' model on each partition of data, then labels an unlabeled public dataset using an aggregation of predictions from each teacher model, and finally trains a model on this dataset. In our work, we focus on model evaluation rather than training, as the model evaluation task is quite challenging in the graph setting. We observe a number of distinct difficulties in applying subsample-and-aggregate to benchmarking fraud detectors on graphs. First, sub-sampling a \emph{graph} often introduces significant bias to the estimates of graph statistics on each subset, which changes assessments of the best choice of the number of partitions to use. Second, in fraud detection problems there are often very small numbers of fraudsters. Hence, sub-sampling these fraudsters can lead to very few per partition and poor utility. We propose up-sampling fraudulent entities satisfying a relaxed notion of privacy to address this issue. We then perform extensive empirical evaluation to understand how the subsample-and-aggregate framework compares to synthetic data generation algorithms for this problem.

A recent work \cite{kohli2023_blackboxdp} also considers the problem of running arbitrary code on a private dataset and gives a new mechanism called TAHOE that finds a subset of the data on which the script is ``stable.'' They compare against subsample-and-aggregate on simple queries like producing a histogram of the data and find their method is competitive in accuracy. However, TAHOE is computationally expensive and can only be run efficiently when the dataset can be expressed as a histogram of finitely many values, which is inapplicable to graph data.

There is a long and rich line of work in differentially private analysis of graph data. We discuss the literature on generating synthetic graphs in more detail in Section~\ref{sec:synth} where we detail our choice of synthetic graph algorithms to benchmark. These synthetic graph algorithms require estimating statistics of the graph (like degree distribution or number of triangles) under node differential privacy. This introduces new challenges as the prior works on synthetic data generation use the weaker notion of edge-DP to estimate statistics. In our work, we generically transform edge-DP estimation into (reasonably accurate) node-DP estimation using the idea of smoothly projecting a graph to the space of limited-degree graphs from \cite{blocki2012_nodedp, kasiviswanathan2013_nodedp}. There may be additional improvements in applying existing synthetic data generation methods under the node-DP privacy regime by applying more tailored estimation procedures for specific graph statistics.

\section{Subsample-and-Aggregate}
\label{sec:subsamp}
The first approach we consider in privatizing fraud benchmarking is the subsample-and-aggregate framework \cite{nissim_smooth_sensitvity_07}. Recall from Section~\ref{sec:graph_challenges} that a key challenge of releasing a DP estimate of the AUC score of a fraud detector on a graph is that this query has global sensitivity of $1$, equal to the range of the AUC score. Subsample-and-aggregate forces low sensitivity of the query by partitioning the dataset into $\npartition$ disjoint sets and estimating AUC on each partition.

Our algorithm follows the template described above for benign vertices, that is, we partition the benign vertices into $\npartition$ disjoint sets of equal size. However, in fraud graphs, there are often very few fraudulent vertices. For example, in the Elliptic Bitcoin financial fraud dataset \cite{weber2019_elliptic} there are only $11$ fraudsters out of over $6,000$ vertices. Partitioning these fraud vertices into a reasonable number of partitions to achieve low sensitivity (say $\npartition \geq 5$) would destroy any structure of the sub-graph of fraud vertices. 

We therefore modify typical subsample-and-aggregate for the fraudulent vertices by allowing \emph{duplication} of fraudsters across partitions. For each partition, we sample a subset of fraudulent vertices, where the rate of sub-sampling is controlled by a parameter $\subsamplerate$. We term this instance of subsample-and-aggregate as Partition, Duplicate, and Aggregate (PDA), described in Algorithm~\ref{alg:pda}. Note that taking $\subsamplerate = 1$ results in duplicating all fraud vertices in each partition, while taking $\subsamplerate = \frac{1}{\npartition}$ is similar to typical subsample-and-aggregate, but with the difference that fraudulent vertices may be sampled into multiple partitions.

\begin{algorithm}[ht]
\caption{Partition, Duplicate, and Aggregate}
\label{alg:pda}
   {\bfseries Parameters:} privacy parameter $\eps > 0$, number of partitions $\npartition$, fraud sub-sampling rate $\subsamplerate$.
   
    {\bfseries Inputs:} fraud detector $\fraudAlgo$, accuracy statistic $\accuracy$ with global sensitivity $\sensitivity$,  fraud vertices $\fraudNodes$, benign vertices $\benignNodes$, graph $\graph$ on vertex set $\benignNodes \cup \fraudNodes$.

    \begin{itemize}[leftmargin=*]
        \item Randomly partition non-fraud nodes $\fraudNodes$ into $\npartition$ equally size sets $\benignNodes^{(1)},\ldots,\benignNodes^{(\npartition)}$.\
        \item Randomly sample $\npartition$ sets of fraud nodes $\fraudNodes^{(1)},\ldots,\fraudNodes^{(\npartition)}$ where each  $\fraudNodes^{(i)}$ is sampled independently uniformly from all sub-sets of $\fraudNodes$ of size $\subsamplerate \cdot |\fraudNodes|$.
        \item Let $\graph_1, \cdots, \graph_\npartition$ be sub-graphs of $G$ on vertices $(\benignNodes^{(1)} \cup \fraudNodes^{(1)}),\ldots,(\benignNodes^{(\npartition)} \cup \fraudNodes^{(\npartition)})$.
        \item Release $Z + \frac{1}{\npartition} \sum_{i=1}^{\npartition} \accuracy\left(\fraudAlgo, \graph_i\right)$ where $Z \sim \text{Laplace}(\sensitivity/(\npartition \eps))$.
        
    \end{itemize}
\end{algorithm}

It is straightforward to prove that Algorithm~\ref{alg:pda} guarantees differential privacy:

\begin{proposition}
    For any choice of sub-sampling rate $\rho \in (0,1)$, number of partitions $\npartition > 1$ and privacy parameter $\eps > 0$, Algorithm~\ref{alg:pda} guarantees $\eps$-Protected Differential Privacy (Definition~\ref{defn:protected_dp}).
\end{proposition}

The proof follows from a standard proof  of privacy for subsample-and-aggregate: changing the data of any benign vertex impacts at most $1$ of the $\npartition$ partitions between any two neighboring graphs, and the accuracy score on this partition can change by at most $1$ since $\accuracy$ has global sensitivity (Definition~\ref{defn:sensitivity}) of $1$. Hence, the mean across partitions has global sensitivity of $\frac{1}{\npartition}$ and privacy follows from the Laplace mechanism (Definition~\ref{defn:laplace}). We note that in practice, the subsample-and-aggregate framework does not introduce substantial computational overhead.

 We further develop intuition around the error of partition, duplicate, and aggregate as a function of number of partitions and subsample-rate via a simple example in Appendix~\ref{app:bias_variance}.

\subsection{Runtime}
\label{sec:computation}

We observe in experiments that subsample-and-aggregate often significantly speeds up evaluation of fraud detectors. In this appendix, we give some theoretical intuition for why this may be. Algorithm~\ref{alg:pda} requires running the same fraud detector on $\npartition$ partitions of the dataset, each of size roughly $\frac{1}{\npartition}$ the original number of vertices in the dataset (in fact, slightly larger due to duplication of fraud vertices). In many cases, the runtime of a fraud detector actually decreases by a factor of more than $\frac{1}{\npartition}$ per partition. This can happen for two reasons. First, many graph algorithms are polynomial in the number of vertices in the graph (for example, an algorithm that cubes the adjacency matrix to compute number of triangles per vertex). Hence, the partitioning gives a polynomial $\frac{1}{k}$ improvement in total computational cost. Second, partitioning can only decrease the total number of edges across all partitions since no edge can be duplicated in multiple partitions. Therefore, any algorithm with runtime dependent on number of edges is faster when run on all the partitioned graphs rather than the original graph. In addition, Algorithm~\ref{alg:pda} can be parallelized by running the fraud detector on each partition separately, which may make it easier for the benchmarking server to efficiently execute submitted fraud detectors in practice.  

\subsection{Running Multiple Benchmarks}
\label{sec:composition}

Algorithm~\ref{alg:pda} provides a method for one-shot release of the AUC score of a single fraud detector. In order to apply it to full leaderboard release, we can invoke composition (Theorem \ref{thm:composition}) and subdivide the privacy budget among many fraud detectors. For example, if we have $10$ algorithms to benchmark, we run each with privacy budget of $\eps / 10$ per fraud detector. As it is harder to provide good utility for smaller $\eps$,  we expect our accuracy of estimation to degrade in the number of detectors benchmarked.

In many real-world settings it is useful to release only the best or the top-$m$ fraud detectors, for example when running a competition. In the case of top-$1$ release, we can use the \emph{Report Noisy Arg Max} mechanism \cite{dwork_roth_textbook}. This mechanism adds Laplace noise to any (finite) number of queries as per the Laplace mechanism, but then only releases the name of the query with the largest (noisy) value. Rather than paying composition cost that grows in the number of queries, this procedure is $\eps$-DP. In our case, then, we can apply Algorithm~\ref{alg:pda} to arbitrarily many fraud detectors and then at the end only publicly release the name of the detector with the highest noisy AUC score. This guarantees $\eps$-DP when each run of Algorithm~\ref{alg:pda} is run using privacy parameter $\eps$. While we do not experiment with releasing the top-$m$ fraud detectors, recent work \cite{qiao21_top_k} shows that releasing the top-$m$ fraud detectors ranked by noisy AUC (among a larger set of fraud detectors) only incurs total privacy loss of $m\eps$.
\section{Synthetic Data Generation}
\label{sec:synth}

In this section, we describe our choice of synthetic graph data generation algorithms to benchmark; the surveys \cite{hu2023_survey, li2021_survey} provide a useful overview of such algorithms. Many methods do not handle \emph{labeled} vertices. Such methods cannot be applied to our problem, as synthetic data for fraud detection benchmarking needs to  differentiate between fraudulent and benign vertices. Additionally, most existing work focuses on satisfying the weaker notion of edge-level DP, while we wish to satisfy node-level DP. Therefore, we focus on the following $3$ methods that all handle labelled vertices and are amenable to transformation into a node-level DP algorithm:

\begin{enumerate}[leftmargin=*]
    \item \emph{Stochastic block model (SBM)}: Estimate a stochastic block model with two communities (fraud and non-fraud) and sample a graph based on the SBM parameters. For a fixed number of benign and fraud vertices, the stochastic block model has three parameters $\pfraud, \pbenign, \pcross$. Each edge in the graph is sampled independently at random with probability $\pfraud$ if both of its endpoints are fraudulent, $\pbenign$ if both are benign, and $\pcross$ if one is fraudulent and the other is benign.
    \item \emph{Attributed social graph} (ASG):~\cite{jorgensen2016_asg}: Estimate the connection probabilities with and between fraud and non-fraud vertices (as in the SBM), but additionally estimate number of triangles in the graph and the degree sequence of the graph. Then, sample a graph that matches these noisy statistics. We run two versions of this method, with and without the triangle statistic.
    \item \emph{Top-m-filter}~\cite{nguyen2015_topm}: Directly perturb the adjacency matrix of the graph. In particular, flip each edge in the graph and then perform a filtering step to remove edges to match a noisy estimate of total number of edges.
\end{enumerate}

In general, synthetic data methods first compute graph statistics under differential privacy, which provide a succinct representation of the graph, and then generate the synthetic graph based on these (noisy) statistics. More expressive graph models may better represent the graph structure, but tend to require the estimation of noisier sufficient statistics due to differential privacy. We choose methods that lie along this spectrum of model complexity. 
On one end, the \emph{SBM} represents a simple model of the graph, with statistics that can be accurately estimated under differentially private. On the other end, \emph{Top-m-filter} attempts to release the full adjacency matrix, which requires large relative noise addition to each entry to guarantee privacy, but best represents the graph structure if privacy were not a concern. 

 We briefly discuss other popular synthetic graph models considered in the literature. One approach uses exponential random graph models (ERGMs) to model vertex-labelled graph data \cite{karwa2015_ergm, liu2020_ergm}. These methods are difficult to scale to graphs of more than a few hundred vertices, and prior empirical evaluations are limited to graphs of this size. Hence, they are not applicable to the types of fraud graphs we consider which are larger by an order of magnitude. Recent work \cite{yoon2023_graph, zahirnia_2024neural} has considered using Graph Neural Networks (GNNs) to generate synthetic graph data from graph statistics. They find that directly using DP-SGD (stochastic gradient descent) to train the GNN leads to poor utility. However, it is possible to obtain useful synthetic data by computing vectors of vertex-level graph statistics (like histograms of triangles and $2$-paths) under edge-level DP. Estimating these sub-graph histogram statistics under \emph{node-level DP} requires much larger noise addition. For instance, even assuming that a graph has no vertices of degree greater than $\degreeThreshold$, the triangle histogram has sensitivity of greater than $\degreeThreshold^2$, while we would expect most vertices to participate in far fewer than $\degreeThreshold$ triangles. Because we would need to add much more noise, than the experiments from this work (which just add noise scaled to $\frac{1}{\eps}$ to the statistics) we do not focus on these methods in this work. Finally, some methods incorporate a community detection step \cite{chen2019_asg, yuan2023_privgraph} that first clusters vertices of the graph and then estimates connection probability parameters between these clusters to incorporate into the graph generative model. It is unclear how to make this clustering step satisfy node-level DP with reasonable utility.

\subsubsection*{Guaranteeing Node-Level Differential Privacy.} 
The algorithms we consider were designed to provide edge-level differential privacy. In privately computing sufficient statistics of the graph, these algorithms add Laplace noise proportional to the worst-case sensitivity of a statistic to the change of a single edge in a graph. In order to guarantee node-level privacy in this noise addition step, we use the idea of projecting the graph to the space of graphs with bounded maximum degree from \cite{blocki2012_nodedp, kasiviswanathan2013_nodedp} and then adding noise proportional to this ``restricted sensitivity.'' For a given graph $\graph$, choice of truncation threshold $\degreeThreshold$, and graph statistic $g$, the full workflow is:
\begin{enumerate}[leftmargin=*]
    \item (Naive truncation). Truncate graph $\graph$ by removing all vertices with degree above $D$.
    \item Estimate the ``smooth sensitivity'' $S$ of the naive truncation operation per \cite{kasiviswanathan2013_nodedp}.\footnote{From \cite{kasiviswanathan2013_nodedp}, Proposition~6.1 we can compute the smooth sensitivity  $S^{\beta}_{trunc}(\graph, \degreeThreshold)$ of the truncation operation as follows. Let $N_t(\graph, \degreeThreshold)$ denote the number of benign vertices with degree in range $[\degreeThreshold - t, \degreeThreshold + t + 1]$ and $C_t(\graph, \degreeThreshold) = 1 + t + N_t(\graph, \degreeThreshold)$. Then, $S^{\beta}_{trunc}(\graph, \degreeThreshold) = \max_{t \geq 0} e^{-\beta t} C_t(\graph, \degreeThreshold)$. }
    \item Add Laplace noise with scale proportional to $S \cdot RS_\degreeThreshold(g)$ where $RS_\degreeThreshold(g)$ represents the ``restricted sensitivity'' of $g$ on graphs of max degree $D$, that is the maximum change in $g$ between any two node-adjacent graphs of max degree $\degreeThreshold$.
\end{enumerate}

We summarize the framework for node-private synthetic data release in Algorithm~\ref{alg:synthetic}. Since the max degree and average degree of the fraud graphs used (see Table~\ref{tab:test_datasets_summary}) tends to be much smaller than the number of vertices in the graph, the restricted sensitivity tends to be much lower than the global sensitivity. For example, the global sensitivity of the number of edges in the graph is $\nbenign-1$, while the restricted sensitivity at threshold $\degreeThreshold$ is only $\degreeThreshold$. We test synthetic data generation algorithms for a variety of choices of threshold.

Note that using this method with Laplace noise actually guarantees the relaxation of $(\epsilon, \delta)$-differential privacy due to the use of ``smooth sensitivity''~\cite{nissim_smooth_sensitvity_07}.  We fix $\delta$ to $10^{-8}$ for all experiments on synthetic data methods. Additionally, to provide a fair comparison against our subsample-and-aggregate method which relaxes privacy for fraudulent vertices, we compute statistics that rely only on the fraudulent nodes without noise.

\begin{algorithm}[ht]
\caption{Framework for Node-Private Synthetic Data Release}
\label{alg:synthetic}
{\bfseries Parameters:} privacy parameters $\eps > 0, \delta \in (0,1)$, degree threshold $\degreeThreshold$

{\bfseries Inputs:} fraud vertices $\fraudNodes$, benign vertices $\benignNodes$, graph $\graph$ on vertex set $\benignNodes \cup \fraudNodes$, vector of sufficient statistics to compute $\sufficientStats(\graph)$ with restricted sensitivity $\Delta_\degreeThreshold$.
\begin{itemize}[leftmargin=*]
    \item Remove all benign vertices from $\graph$ with degree greater than $\degreeThreshold$.
    \item Compute the $\beta$-smooth sensitivity $S^{\beta}_{trunc}(\graph, \degreeThreshold)$ of the truncation operation on $\graph$, where $\beta = - \frac{2 \eps}{\log(1/2\delta)}$.
    \item Release $\widetilde{\sufficientStats}(\graph) = \sufficientStats(\graph) + Z$ where $Z \sim \text{Laplace}(2 S^{\beta}_{trunc}(\graph, \degreeThreshold) \cdot  \Delta_\degreeThreshold / \eps)$.
    \item Sample output synthetic graph $\widetilde{\graph}$ based on $\widetilde{g}(\graph)$.
\end{itemize}
\end{algorithm}

\section{Experimental Setup}
\label{sec:methods}

In the following section, we describe our datasets, fraud detectors, and metrics used in empirical evaluations.

\subsection{Datasets}

\begin{table}[ht]
    \centering
    \begin{tabular}{lrrrrr}
    \toprule
     & Yelp  & Amazon & Peer Review & Elliptic \\
    \midrule
    Vertices & 11,473 & 11,944  & 2,483 & 6,621 \\
    Edge Density (\%) & 0.41 & 6.17 & 0.77 & 0.04 \\
    Num Fraud & 1,657 & 821 & 22 & 11 \\
    Max Degree & 236 & 6,991& 255 & 47 \\
    Mean Degree & 47.45 & 736.50 & 19.12 & 2.51 \\    \bottomrule
    \end{tabular}
    \caption{Graph test datasets}
    \label{tab:test_datasets_summary}
\end{table}

We test methods for fraud benchmarking on 4 datasets representing a variety of domains and graph structures. All graphs are undirected unipartite graphs. In \emph{Yelp} \cite{dou2020_yelpamazon} and \emph{Amazon} \cite{dou2020_yelpamazon} each vertex represents a reviewer with edges denoting common reviews on the products/restaurants and fraudulent reviewers represent spammers and low-rated reviewers respectively. \emph{Peer Review} consists of paper reviewers at a computer science conference with edges denoting mutual bids on each other's papers \cite{wu_robust_peer_review_assignment}. Following \cite{jecmen2024_detection} we inject a clique of $22$ fraudulent reviewers with edge density of $0.8$ among these reviewers into the graph, which corresponds to the smallest injected clique that was possible to detect in prior work. Finally, in \emph{Elliptic} \cite{weber2019_elliptic} each vertex in the graph represents a transaction from the Bitcoin blockchain, an edge represents a flow of Bitcoins between one transaction and the other, and fraudulent nodes are illicit transactions. We take a single time-step from the entire Elliptic graph (summary statistics in Table~\ref{tab:test_datasets_summary}).

We run analyses of subsample-and-aggregate and synthetic data algorithms on validation datasets to understand settings of hyperparameters before comparing these methods against each other. We use four validation datasets. For Yelp, we use a random split of the vertices with ~11k vertices in the test set and ~11k in the validation set. For Elliptic, we use different disjoint time periods for validation and test. For Amazon and Peer Review, there are not standard train-test splits used in past work. We therefore use the entire graph for evaluation, and generate validation graphs to set hyperparameters by estimating parameters of a stochastic block model (SBM) and sampling from this model.

\subsection{Fraud Detectors}

We evaluate 10 simple fraud detectors that do not require learning, and 3 detectors that learn on a subset of the benchmark data.
Specifically we evaluate the following fraud detectors:
\begin{itemize}[leftmargin=*]
    \item \emph{(Negative) Degree~\cite{jecmen_dataset}}: rank by the degree of each vertex.
    \item \emph{(Negative) Clustering Coefficient}: rank by the clustering coefficient of each vertex, inspired by \cite{akoglu2010oddball}.
    \item \emph{SVD Error~\cite{jecmen_dataset}}: take the singular value decomposition of the adjacency matrix to obtain a low rank approximation (for specified rank $r$). Then, rank each vertex by reconstruction error (aggregating over edges by taking either the sum or the max over edges). We use $r = 10$ for the sum  and $r = 50$ for the max, chosen to maximize average AUC across all datasets in a grid search.
    \item \emph{Community Detection}: run Leiden community detection \cite{traag2019_louvain} to place cluster vertices in a cluster and rank by cluster size (with larger clusters less likely to be fraudulent).
    \item \emph{Aggregations:} take weighted averages of the (normalized) scores or maximum scores obtained from subsets of the prior methods.
    \item \emph{GraphSAGE (SAGE)~\cite{graphsage}}: learns a function that aggregates embeddings from neighboring nodes to learn a node embedding for each vertex and uses these embeddings to predict fraud labels.    
    \item \emph{Graph Convolutional Network (GCN)~\cite{gcn}}: learns node representations by applying a layer-wise convolutional operation that aggregates and normalizes features from immediate neighbors.
    \item \emph{Graph Attention Network (GAT)~\cite{gat}}: employs attention mechanisms to weight the influence of neighboring nodes during aggregation.
\end{itemize}

The last three detectors involve a learning component; 
we trained each GNN on a random 80\% of vertices and then assessed AUC score on the held-out 20\%.
These algorithms give a wide range of AUC scores on each dataset. For example, on Yelp, GCN performs the best with an AUC score of $0.73$ and SVD Error (Sum) performs poorly with an AUC score of $0.34$. In contrast, on Peer Review, SVD Error (Sum) performs the best with an AUC score of $0.88$, while Neg Degree has very bad performance with AUC of $0.12$.

\subsection{Measuring Utility}

We consider three metrics to compare utility across methods. Each metric corresponds to one of the release modes for the benchmarking server: one-shot, full leaderboard and top-1 release. Let $\{\fraudAlgo_i\}_{i=1}^\numalgos$ denote a set of $\numalgos$ fraud detectors to benchmark on graph $\graph$, $\auc(\fraudAlgo_i, \graph)$ denote the true AUC score for fraud detector $i$ on $\graph$ and $\aucpriv(\fraudAlgo_i, \graph)$ denote the noisy DP estimate of the AUC score. For the one-shot release, where we wish to release AUC for a single fraud detector, we calculate \emph{$L1$ error}: $|\auc(\fraudAlgo_i, \graph) - \aucpriv(\fraudAlgo_i, \graph)|$.

When evaluating top-$1$ release of the best fraud detector among a set of fraud detectors we measure utility by the distance between the true AUC of the true best fraud detector (computed without any privacy) and the true AUC of the released best fraud detector. That is, we define top-1 error as:
\begin{align*}
    \auc(\fraudAlgo_{top}, \graph) - \auc(\fraudAlgo_{top'}, \graph) & \; \text{where } top = \sigma^{-1}(1), top' = \widetilde{\sigma}^{-1}(1).
\end{align*}
For the full leaderboard setting we use the weighted Kendall-Tau distance between rankings~\cite{kumar_kendalltau}. We discuss this metric and provide results in this setting in Appendix~\ref{app:full_leaderboard}.

In addition to using the AUC score as an accuracy metric, we evaluate the F1 score of fraud detectors, another popular measure of fraud detector accuracy. We find similar results to that of AUC score, but F1 score tends to be even more difficult to release accurately. We present these additional results in the arXiv version of this paper.

\section{Experimental Results}
\label{sec:results}

In this section we provide results of our comparison of Subsample-and-Aggregate and Synthetic Graph Generation~(\ref{sec:all_experiments}), and experiments to understand trade-offs between distorting graph structure and adding noise to preserve privacy for Subsample-and-Aggregate~(\ref{sec:pda_experiments}) and Synthetic Graph Generation~(\ref{sec:synth_experiments}). 

\subsection{Comparison of Algorithms}
\label{sec:all_experiments}

We benchmark subsample-and-aggregate against synthetic data algorithms for the concrete task of releasing the best fraud detectors among a set of fraud detectors. We choose parameters of subsample-and-aggregate (number of partitions and sub-sampling rate) based on the best parameters for each dataset in releasing a ranking of all fraud detectors on the validation dataset. This follows prior work \cite{mohan2012_gupt}, which assumes access to public datasets from which one could estimate subsample-and-aggregate hyperparameters.

\begin{figure}
    \centering
   \includegraphics[width=0.4\linewidth]{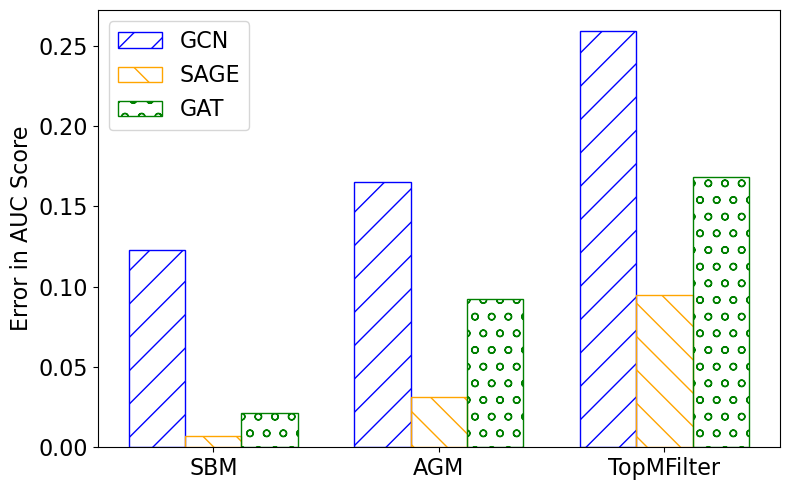}
    \caption{Error in AUC score for GNN-based fraud detectors on Yelp data using synthetic graphs with $\eps=5.0$.}
    \label{fig:synth_graph_gnns}
\end{figure}

In Figure~\ref{fig:head_to_head_primary}, we show results of the DP benchmarking methods for top-$1$ release. We decompose the error into \emph{inductive bias} from how a method distorts graph structure, and error from the addition of \emph{privacy-preserving random noise}. To visualize this, we plot each method without privacy-preserving noise on the x-axis and with noise needed to preserve privacy on the y-axis. Specifically, for Non-Private subsample-and-aggregate we only apply the graph partitioning and do not add Laplace noise to the AUC score. For non-private synthetic data methods we compute sufficient statistics for each method without any noise addition and then generate a graph using those sufficient statistics. Among synthetic data methods, Topmfilter has no error without privacy as it releases the full adjacency matrix, while SBM and AGM introduce error even without privacy. However, after adding noise needed for privacy, SBM performs the best among synthetic data methods. Perhaps surprisingly, this is true even for GNN-based fraud detectors as shown in Figure~\ref{fig:synth_graph_gnns}. Subsample-and-aggregate distorts graph structure extensively, even with only $5$ partitions, resulting in high error without privacy. We provide additional results for larger privacy budget and other datasets in Appendix~\ref{app:top1}, but the general trends are similar.

\subsection{Subsample-and-Aggregate}
\label{sec:pda_experiments}

\begin{figure}[ht]
    \centering
    \includegraphics[width=0.8\linewidth]{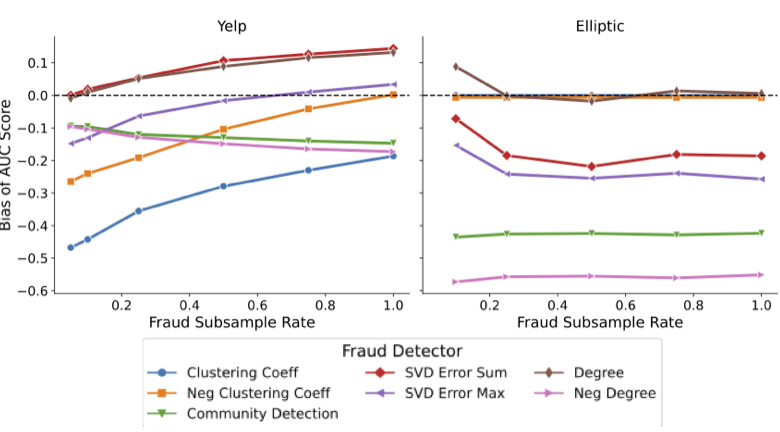}
    \caption{Bias to the AUC score introduced by subsample-and-aggregate for each fraud detector varying the fraud subsample rate $(\subsamplerate)$ while fixing number of partitions $\npartition = 20$. Subsample-and-aggregate introduces extensive bias to all fraud detectors, with the sign and magnitude of the bias varying widely across fraud detectors.}
    \label{fig:bias_vs_subrate}
\end{figure}

\begin{figure}[ht]
\centering
        \includegraphics[width=\linewidth]{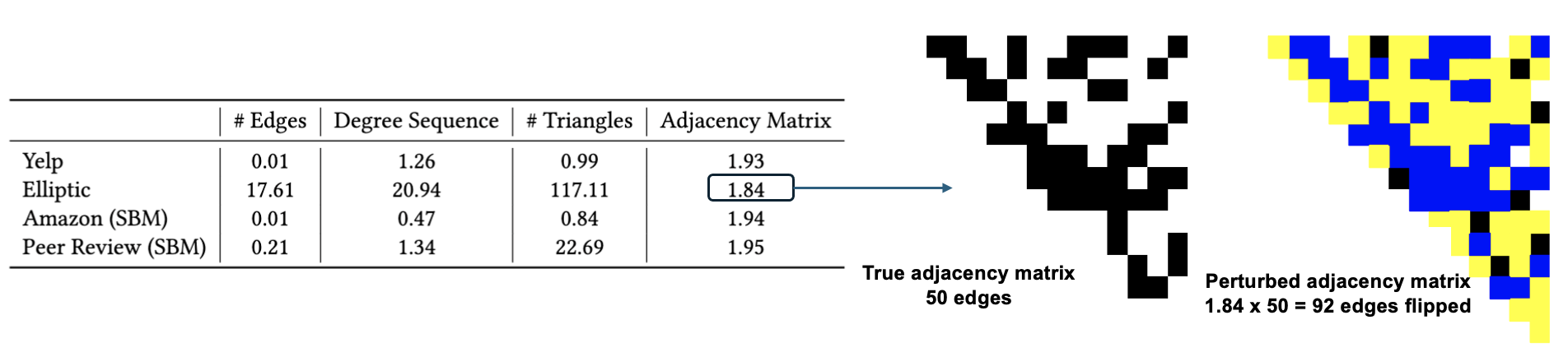}
        
      \caption{Normalized mean absolute error (MAE) introduced to each of the synthetic graph sufficient statistics. We fix $\eps = 5.0$ and the degree cutoff to $1\times$  the graph's max degree. It is possible to estimate SBM parameters accurately, while other parameters have large noise addition. We give an example of what relative error of 1.84 means for the adjacency matrix on the right, where an adjacency matrix with 50 edges has 92 edges flipped: 42 removed (yellow) and 50 added (blue.)}
      \label{tab:suff_statistics}
\end{figure}

In experiments on four validation datasets, we seek to understand how the parameters of the algorithm---number of partitions $\npartition$ and rate of sub-sampling fraud in each partition $\subsamplerate$---impact the bias, variance from Laplace noise addition and overall distortion of fraud detector rankings.

On each dataset we run Algorithm~\ref{alg:pda} for $10$ trials for each choice of parameters $\npartition, \subsamplerate$ and $\eps$. We show per-dataset results on the Yelp and Elliptic validation datasets in this section. We provide additional results on Amazon and Peer Review in Appendix~\ref{app:subsample_aggregate}.
    
 In general, we find that partitioning the graph into random sub-graphs introduces significant bias to estimates of graph statistics. The sign and magnitude of this bias can differ widely across fraud detectors. In Figure~\ref{fig:bias_vs_subrate}, we show bias per fraud detector fixing the number of partitions at $\npartition = 20$ and varying the fraud sub-sampling rate. We find that it is not always possible to achieve zero bias for a given fraud detector for a given number of partitions $\npartition = 20$. For instance, the clustering coefficient detector has negative bias on the Yelp dataset at all values of $\subsamplerate$. This makes sense as removing benign vertices from the graph may change the distribution of fraud detection scores for benign vertices such that it is not possible to recover a similar distribution at any rate of sub-sampling fraudulent vertices. We additionally find, as expected, that the magnitude of bias increases with the number of partitions $(\npartition)$ although the sign of the bias differs across fraud detectors. We plot bias as a function of number of partitions in Figure~\ref{fig:bias_vs_k} of Appendix~\ref{app:subsample_aggregate}. These results explain the poor performance of subsample-and-aggregate in Figure~\ref{fig:head_to_head_primary}, as subsampling tends to distort graph structure extensively, biasing different fraud detectors in different ways thereby undermining the utility of the ranking of fraud detectors.

\subsection{Synthetic Graph Generation}
\label{sec:synth_experiments}

In our experiments we aim to isolate error introduced due to choice of graph model and noisy estimation of sufficient statistics. For each synthetic data generation algorithm, we generate $10$ synthetic data sets. For each synthetic graph method, we subdivide the privacy budget evenly between the different parameters to estimate.  We note that it may be possible to better distribute privacy budget between different statistics, which is an interesting area for future investigation.
 We test degree truncation thresholds as a function of the max degree of each graph, so $1.0$ is a threshold exactly equal to the maximum degree benign vertex in a graph while $0.5$ removes all nodes with degree $>0.5$ times the max degree. In this section, we report results with threshold of $1$ and give additional results for $0.5$ in Appendix~\ref{app:synthetic_data}.

We find that outside of the SBM, it is necessary to introduce large distortion to the sufficient statistics of each graph model in order to preserve privacy, as shown in Figure~\ref{tab:suff_statistics}. We show the proportional change in each noisy sufficient statistic compared to its true value, taking the mean over a vector-valued statistic. On Yelp, Amazon, and Peer Review it is possible to estimate the edge count for the SBM with high accuracy at $\eps = 5.0$, perturbing the edge count by $1\%$ of the total number on Yelp and Amazon.  Elliptic is an extremely sparse graph (0.04\%), so we introduce much larger relative error. For degree sequence and number of triangles, the amount of error is one to two orders of magnitude larger, with error generally at least $50\%$ of the value of the original statistic. Unsurprisingly, the adjacency matrix cannot be accurately estimated under node-DP via direct noise addition. We highlight the amount of noise addition needed to preserve privacy in a simple example of relative error of 1.84 on a 15x15 adjacency matrix, shown in Figure~\ref{tab:suff_statistics}. Even after aggressively truncating high-degree nodes, the addition of DP noise results in flipping the same number of edges as were originally in the adjacency matrix. This large distortion of sufficient statistics explains the poor accuracy of AGM and TopMfilter.

\section{Discussion}
\label{sec:discussion}

In this work we define the novel problem of privately benchmarking fraud detectors on graph-structured data. We benchmark two popular frameworks from the DP literature, subsample-and-aggregate and synthetic data generation. We characterize a trade-off for each method between error arising from bias due to distorting graph structure and error arising from privacy-preserving noise addition. Our results suggest the need to develop methods that trade-off more effectively between graph distortion and noise addition. There are a number of open directions in moving towards this goal:
\begin{enumerate}[leftmargin=*]
    \item \emph{Model / hyper-parameter selection under privacy constraints:} Our experiments suggest that choice of hyperparameters (e.g., number of partitions in subsample-and-aggregate) and more generally method can have a large impact on utility, raising the problem of how to choose the model and hyper-parameters privately.
    \item \emph{General vs. tailored methods of synthetic graph generation:} There are not existing DP synthetic graph algorithms specifically tailored to fraud detection. In our experiments, we find that existing methods introduced significant bias  even without noisy sufficient statistics, suggesting that these models do not capture the structure of graphs needed to model fraudulent behavior.
    \item \emph{Modeling synthetic graph meta-data:} Existing synthetic graph methods try to directly model graph structure. Our experiments demonstrate that this is challenging due to the sensitivity of many graph statistics. We hypothesize that modeling graph meta-data can lead to more effective DP synthetic graph generation methods as meta-data is attributable to one vertex and  can therefore be modeled as tabular data. Then, connectivity between vertices could be estimated using lower sensitivity edge counts between clusters of vertex features as in an SBM.
\end{enumerate}
In conclusion, our work highlights privacy vulnerabilities in benchmarking fraud detectors on private data and explores the challenges in balancing privacy and utility on graph-structured data.

\section*{Acknowledgments}
We thank Patrick Grother and Craig Watson of NIST for  constructive discussion and comments. 
A. Goldberg and G. Fanti acknowledge the Air Force Office of Scientific Research under award
number FA9550-21-1-0090, NSF grant CNS-2148359, the Bill \& Melinda Gates Foundation, Intel, and the Sloan Foundation for their generous support.
A. Goldberg and N. Shah acknowledge the support of NSF grant 2200410 and ONR grant N000142212181. 
S. Wu acknowledges the support of NSF grant 2232693.

\bibliographystyle{alpha}
\bibliography{bibtex}

\appendix
~\\~\\ \noindent{\bf \LARGE Appendices} 

\section{Controlling Bias and Variance in Subsample-and-Aggregate}
\label{app:bias_variance}

Algorithm~\ref{alg:pda} introduces error to the outputted AUC in three ways. First, sub-sampling the graph may introduce bias to the AUC score estimated on each partition; that is, $\E[\auc(\graph_i) - \auc(\graph)] \not= 0$. Second, the algorithm adds Laplace noise to the released statistic, with variance proportional to the inverse of the number of partitions $\npartition$. Finally, estimating on sub-samples of the data may increase the variance of the estimate. 

\paragraph{Tabular data.}
In the special case of benchmarking a fraud detector using tabular data (e.g., only using vertex metadata not graph structure) with full duplication of fraudulent vertices ($\subsamplerate=1$), Algorithm~\ref{alg:pda} introduces error only from noise addition. In particular, changing the value of any one row in a dataset does not change properties of other rows of the dataset. Therefore, on tabular data, partitioning does not change the fraud scores of individual datapoints compared to running the fraud detector on the entire dataset. Then from the definition of the AUC score, the AUC score of sub-partition $i$ is

\[\auc(\fraudAlgo, \graph_i) =  \frac{\npartition}{\nfraud \nbenign} \sum_{v_0 \in \benignNodes^{(i)}} \sum_{v_1 \in \fraudNodes} \ind[\fraudAlgo(\graph, v_1) > \fraudAlgo(\graph, v_0)]\] 

where $\fraudNodes^{(i)} = \fraudNodes$ since we duplicate all fraud vertices in each partition. Then, the mean over all partitions is:

\[
\frac{1}{\npartition} \sum_{i=1}^{k} \auc(\fraudAlgo, \graph_i) \]
\[= \frac{1}{\nfraud\nbenign} \sum_{v_0 \in \benignNodes} \sum_{v_1 \in \fraudNodes} \ind[\fraudAlgo(\graph, v_1) > \fraudAlgo(\graph, v_0)] = \auc(\fraudAlgo, \graph)
\] 

so the mean AUC over partitions exactly recovers the AUC score evaluated on the whole graph. In fact, taking $\npartition = \nbenign$ we add Laplace noise with scale proportional to $\frac{1}{\nbenign}$ in the aggregation step, exactly recovering the Laplace mechanism for a query with global sensitivity of $\frac{1}{\nbenign}$.

\paragraph{Graph data.}
For graph data, partitioning can introduce bias to the estimate of AUC of each partition. The magnitude and direction of the bias may depend on the combination of graph and fraud detector under evaluation. We show this by way of a stylized example. Consider a fraud detector $\fraudAlgo$ which scores each vertex in the graph by its degree. Let graph $\graph$ be a random sample from a very simple stochastic block model (SBM)~\cite{holland_sbm_dfn}. The SBM is defined as follows: fix a number of fraudulent vertices $\nfraud$ and benign vertices $\nbenign$. Then for each pair of fraudulent vertices in the graph, sample an edge between the two independently at random with probability $\pfraud$. For each pair of benign vertices, sample an edge between the two vertices i.i.d. with probability $\pbenign$. Letting the random variable $\degDiff$ be the difference in degree between a random fraudulent vertex and a benign vertex, we are concerned with the expected AUC score of a graph sampled from the SBM model, which is exactly $\E[\ind[\degDiff > 0]] = \Pr[\degDiff > 0]$. 

The expected difference between degree of a fraud vertex and degree of a benign vertex is
\[\E[\degDiff] = (\nfraud - 1) \pfraud - (\nbenign - 1) \pbenign\]
while its standard deviation is given by 
\[\SD[\degDiff] =  \sqrt{(\nfraud - 1) \pfraud (1-\pfraud) + (\nbenign - 1) \pbenign (1-\pbenign)}  \]

For large $\nfraud \pfraud$ and $\nbenign \pbenign$, $\degDiff$ can be approximated by a Normal distribution so we can estimate the expected AUC score as \[\Pr[\degDiff > 0] \approx \Pr\left[Z > -\frac{\E[\degDiff]}{\SD[\degDiff]} \right]\] where $Z$ is a standard normal random variable.
Suppose that $\pfraud > \pbenign$ and $\E[\degDiff] > 0$, so ranking by a degree is a good estimator in that $\Pr[\degDiff > 0] > 0.5$. Now, consider what happens to expected AUC score of a sub-partition, where fraud and benign vertices are sub-sampled at the same rate ($\subsamplerate = \frac{1}{\npartition}$). In this case, $\E[\degDiff]$ decreases by a factor of roughly $\frac{1}{\npartition}$ while $\SD[\degDiff]$ decreases by a factor of $\frac{1}{\sqrt{\npartition}}$ so $-\frac{\diffMean}{\diffSD}$ gets larger, negatively biasing the AUC score downwards. In contrast, consider setting $\subsamplerate = 1$ so all fraud vertices are duplicated. In this case, $\diffMean$ actually increases since we have only down-sampled benign vertices, while the variance decreases, so $-\frac{\diffMean}{\diffSD}$ decreases putting upward bias on the AUC score.

\begin{figure}[ht]
    \centering
    \includegraphics[width=0.5\linewidth]{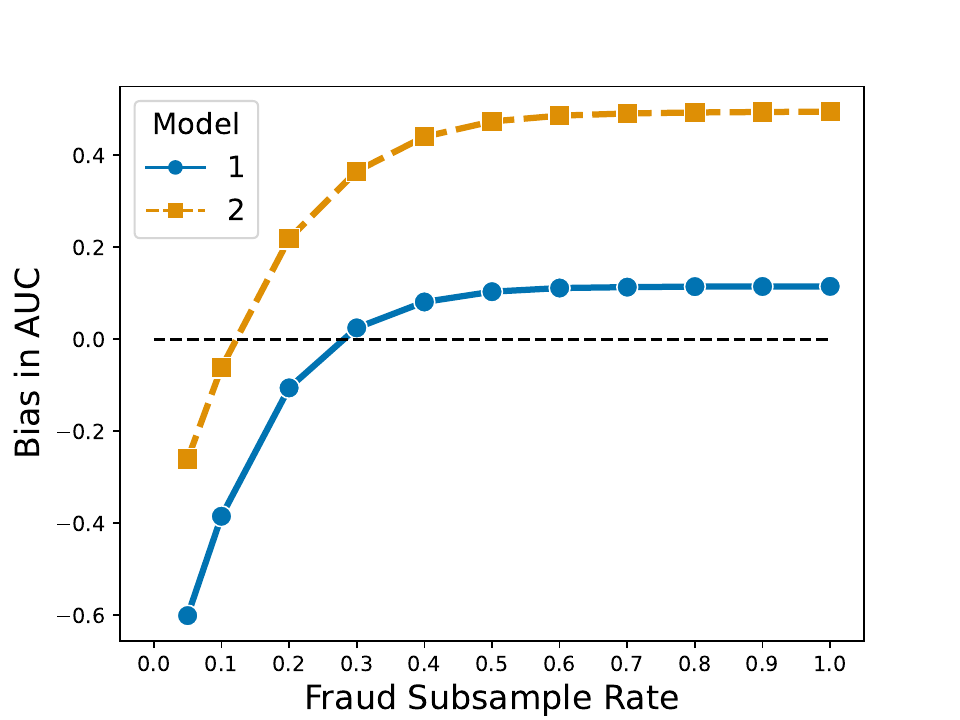}
    \caption{Comparison of the bias in sub-graph AUC score of the ``rank by degree'' detector as a function of fraud subsampling rate $\subsamplerate$ fixing $\npartition = 20$. Results are for two simulated SBM models on $1,000$ benign and $100$ fraud vertices. The bias-minimizing choice of parameter is between $1/\npartition$ and $1$, but quite different on the two datasets.}
    \label{fig:sbm_sims}
\end{figure}

In theory, then, we would like to choose a sub-sampling rate somewhere between $\frac{1}{\npartition}$ and $1$ to minimize this bias. Unfortunately, it is unclear how to set this sub-sampling rate in general. For example, in Figure~\ref{fig:sbm_sims}, we show simulations for simple SBM models on $1,000$ benign and $100$ fraud vertices where Model 1 has $\nbenign \pbenign = 5$, $\nfraud \pfraud = 10$, while Model 2 has $\nbenign \pbenign = 9$, $\nfraud \pfraud = 10$. In both cases, we can observe that taking $\subsamplerate = 1.0$ leads to positive bias while $\subsamplerate = 1 / \npartition$ leads to negative bias. However, the two differ in optimal choice of subsample rate. For Model 1, the best subsample rate for $0$ bias is roughly $0.3$, which gives large positive bias on Model 2. Meanwhile, for Model 2, the best subsample rate for $0$ bias is around $0.1$, which gives significant negative bias on Model 2. It is unclear how to choose the error-minimizing parameters as privately estimating the \emph{error} in AUC requires estimating the AUC, the original estimation problem. In prior work on subsample-and-aggregate building a system named GUPT, the authors advocate for choosing parameters of subsample-and-aggregate based on older (now) public data that is similar to the private dataset \cite{mohan2012_gupt}. However, such data may be difficult to find in the fraud setting. In our work, we empirically evaluate error as a function of $\npartition$, $\subsamplerate$, and $\eps$ on validation datasets in Section~\ref{sec:pda_experiments} and then use the best choice of parameters on test datasets for comparison against synthetic data methods.

\section{Additional Results}
\label{app:additional_results}

In this section we present additional results of our experiments.

\subsection{Comparison of Algorithms, Top-1 Release}
\label{app:top1}

\begin{figure}[ht]
    \centering
    \begin{subfigure}[b]{0.48\linewidth}
        \centering
        \includegraphics[width=\linewidth]{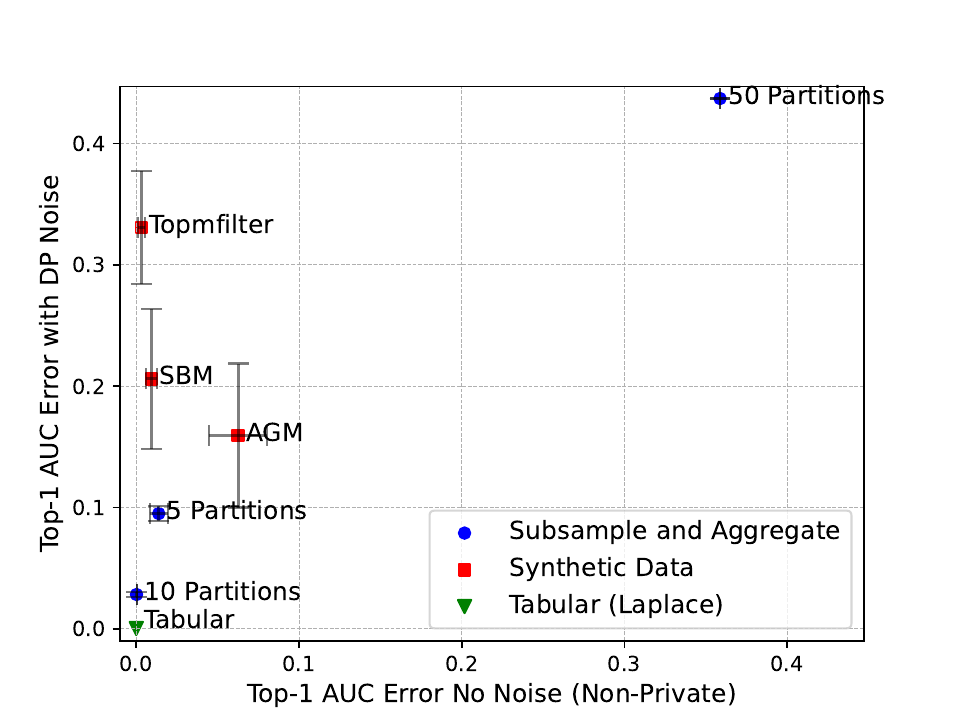}
        \caption{Elliptic Data}
        \label{fig:elliptic_auc_5}
    \end{subfigure}
    \hfill
    \begin{subfigure}[b]{0.48\linewidth}
        \centering
        \includegraphics[width=\linewidth]{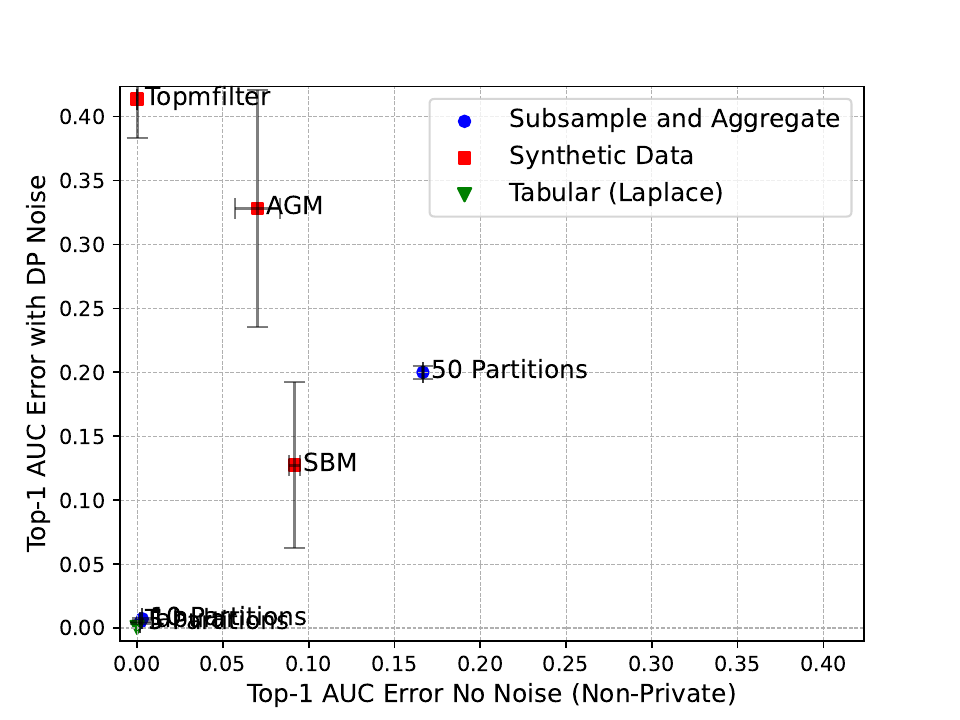}
        \caption{Peer Review Data}
        \label{fig:peer_review_auc_5}
    \end{subfigure}
    \caption{Top-1 AUC, $\eps=5.0$}
    \label{fig:elliptic_peer_review_auc_5}
\end{figure}

\begin{figure}[ht]
    \centering
    \begin{subfigure}[b]{0.48\linewidth}
        \centering
        \includegraphics[width=\linewidth]{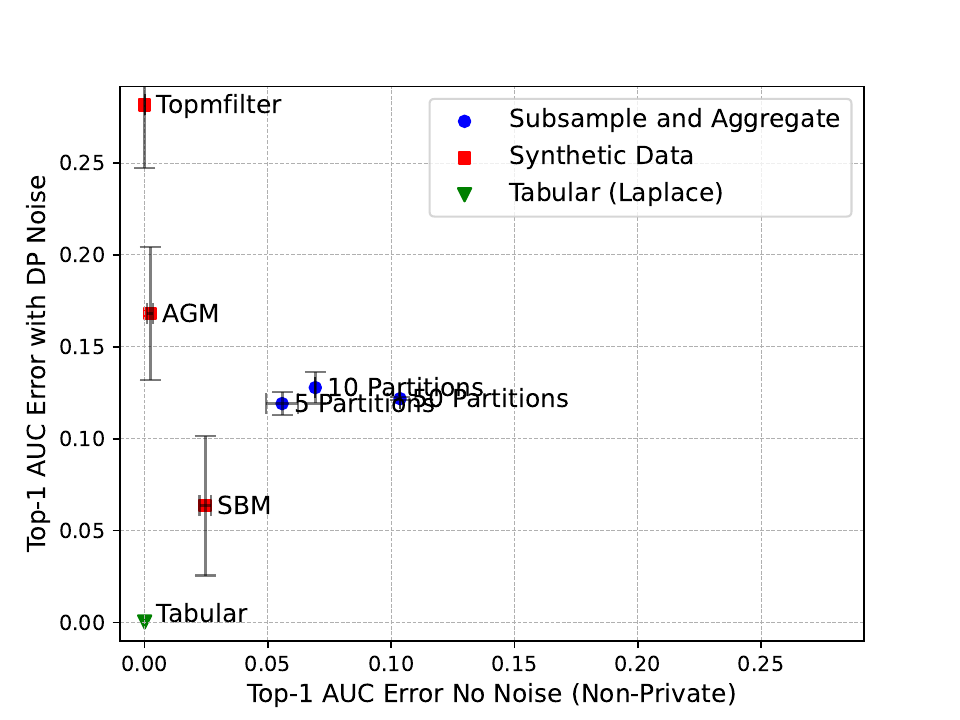}
        \caption{Amazon Data}
        \label{fig:amazon_top1_2}
    \end{subfigure}
    \hfill
    \begin{subfigure}[b]{0.48\linewidth}
        \centering
        \includegraphics[width=\linewidth]{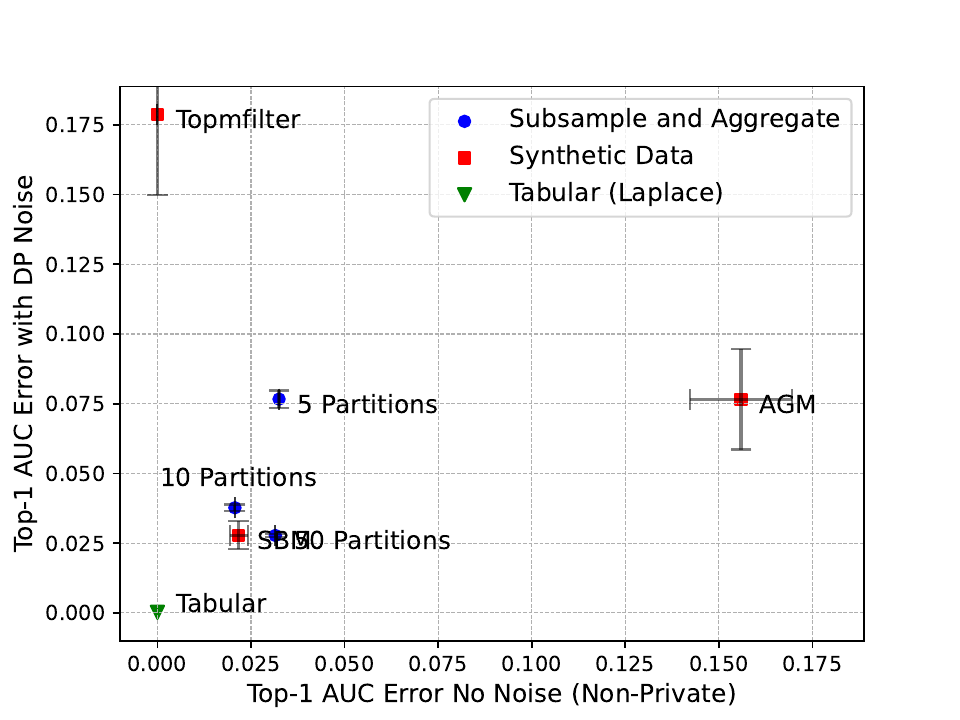}
        \caption{Yelp Data}
        \label{fig:yelp_top1_2}
    \end{subfigure}
    \caption{Top-1 AUC error, $\eps=2.0$}
    \label{fig:amazon_yelp_top1_2}
\end{figure}

\begin{figure}[ht]
    \centering
    \begin{subfigure}[b]{0.48\linewidth}
        \centering
        \includegraphics[width=\linewidth]{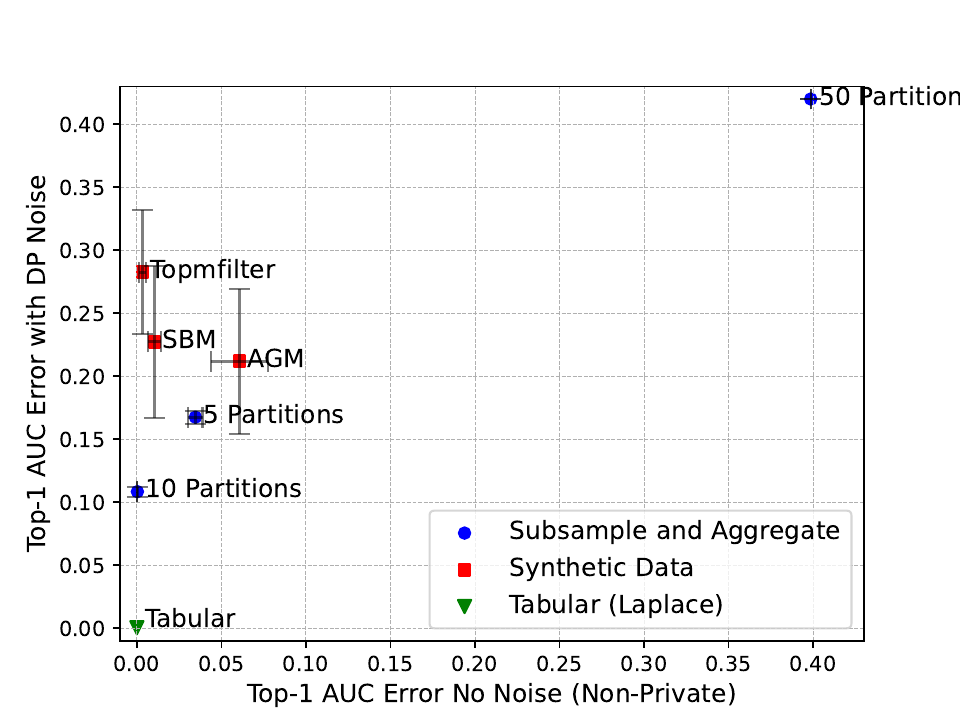}
        \caption{Elliptic Data}
        \label{fig:elliptic_top1_2}
    \end{subfigure}
    \hfill
    \begin{subfigure}[b]{0.48\linewidth}
        \centering
        \includegraphics[width=\linewidth]{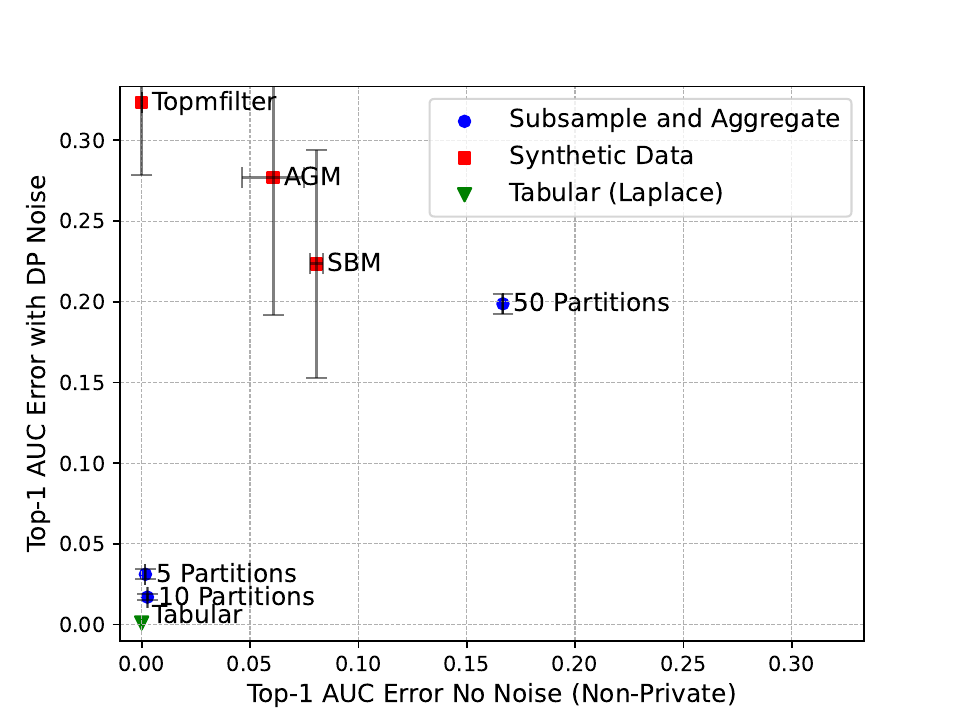}
        \caption{Peer Review Data}
        \label{fig:peer_review_top1_2}
    \end{subfigure}
    \caption{Top-1 AUC error, $\eps=2.0$}
    \label{fig:elliptic_peer_review_top1_2}
\end{figure}

In this section, we provide additional results for the comparison of all algorithms at top-1 release as measured by the error in AUC score of the released best fraud detector vs. the actual best fraud detector. 
In Figures~\ref{fig:elliptic_peer_review_auc_5}, we show the same plot as Figure~\ref{fig:head_to_head_primary} for the Elliptic and Peer Review datasets. We generally observe a similar trend of increasingly complex methods performing worse afer private noise addition than simpler methods. Interestingly, on Elliptic, subsample and aggregate works well with $\eps=5.0$. We note that Elliptic is much sparser than the other graphs, so it may admit different effective algorithms. 
In Figures \ref{fig:amazon_yelp_top1_2} and \ref{fig:elliptic_peer_review_top1_2}, we give results for the stricter privacy budget of $\eps = 2.0$. 

\subsection{Comparison of Algorithms, Full Leaderboard}
\label{app:full_leaderboard}

For the full leaderboard, to capture distance between the true ranking of fraud detectors and the privacy-preserving noisy ranking, we use a similarity-weighted Kendall-Tau distance \cite{kumar_kendalltau}, which counts the number of inversions between two rankings, weighted by the difference in true AUC scores of the swap. Precisely, let  $\sigma(i)$ denote the rank of fraud detector $\fraudAlgo_i$ in the true AUC leaderboard and $\widetilde{\sigma}(i)$ denote the rank of fraud detector $\fraudAlgo_i$ in the noisy AUC leaderboard. Then, the similarity weighted Kendall-Tau distance is given by:
\[ \sum_{(i,j): \sigma(i) < \sigma(j)} \ind[\widetilde{\sigma}(i) > \widetilde{\sigma}(j)] \left(\auc(\fraudAlgo_i, \graph ) - \auc(\fraudAlgo_j, \graph)\right).   \]
As a baseline value for the Kendall-Tau similarity on our set of $10$ fraud detectors on each dataset, we can compute the expected distance between the true leaderboard and a random permutation of the fraud detectors for each dataset. This yields values in the range of $5$ to $8$ for each dataset (which we show as baselines in our results section). 
For further validation of the metric, we consider the distance between rankings on validation and test sets for the Yelp and Elliptic datasets. We find that the distance from test to validation is $0.003$ and $0.021$ respectively reflecting that test and validation sets reliably produce similar leaderboards.

\begin{figure}[ht]
    \centering
    \begin{subfigure}[b]{0.48\linewidth}
        \centering
        \includegraphics[width=\linewidth]{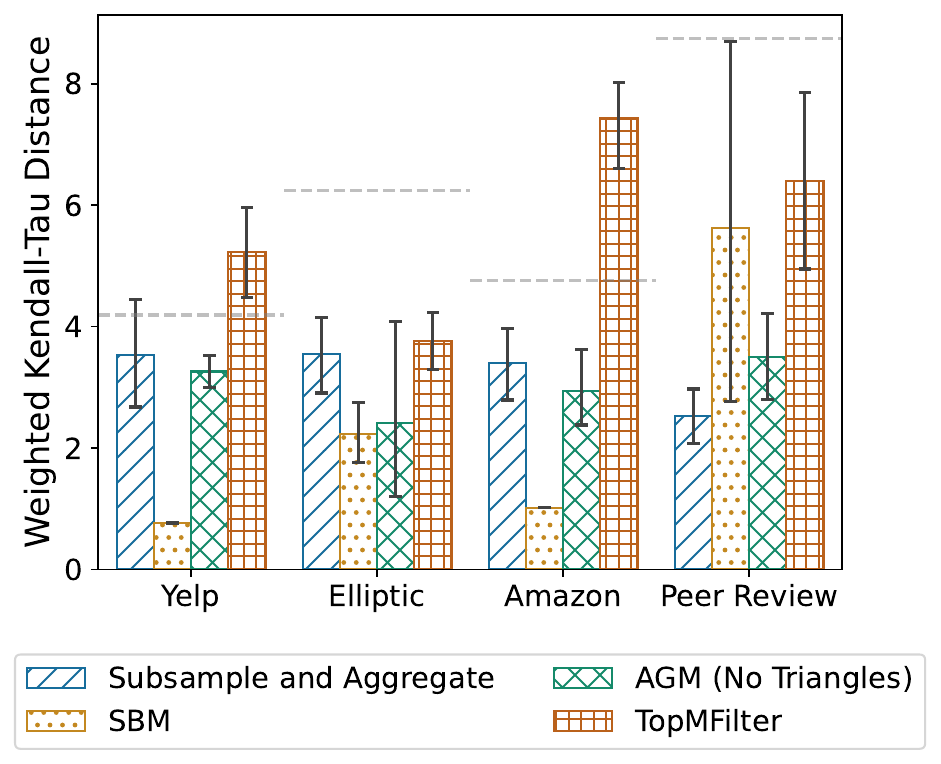}
        \caption{$\eps=5.0$}
        \label{fig:kendall_tau_5}
    \end{subfigure}
    \hfill
    \begin{subfigure}[b]{0.48\linewidth}
        \centering
        \includegraphics[width=\linewidth]{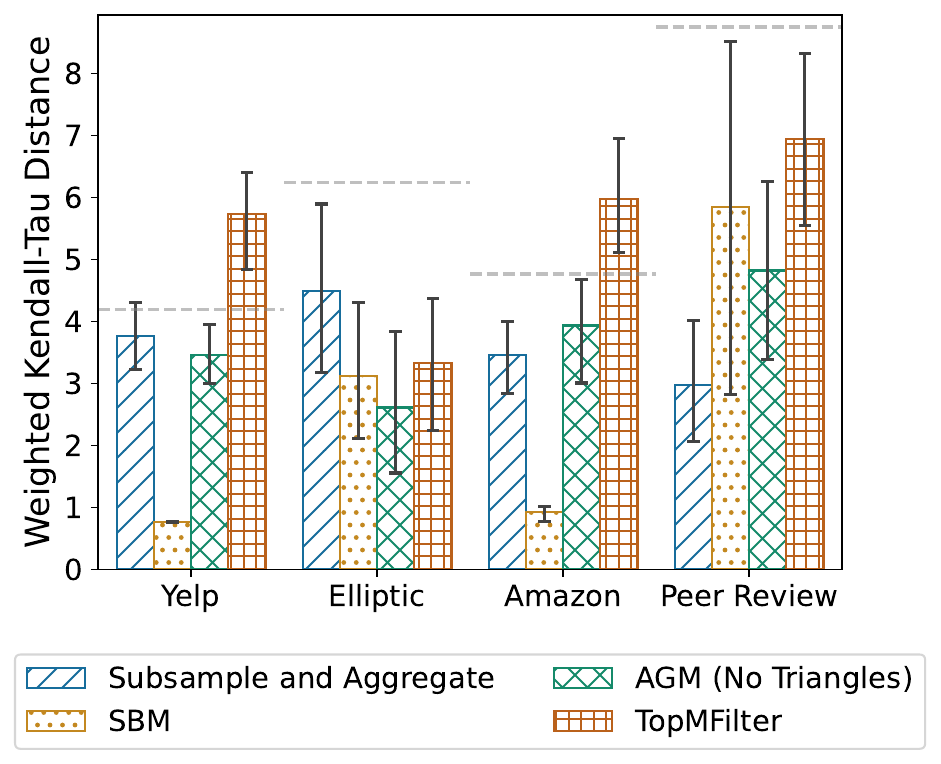}
        \caption{$\eps=2.0$}
        \label{fig:kendall_tau_2}
    \end{subfigure}
    \caption{Head-to-head comparison of DP benchmarking methods. Dashed lines show expected Kendall-Tau distance of a random permutation. Error bars show standard errors over 10 trials. SBM and subsample-and-aggregate are the most competitive approaches, though neither uniformly outperforms the other.}
    \label{fig:kendall_tau_comparison}
\end{figure}

In Figure~\ref{fig:kendall_tau_comparison}, we show the Kendall-Tau distance on each dataset for the best choice of parameters with privacy budgets of $\eps=5.0$ and $\eps=2.0$ respectively. Subsample-and-aggregate is generally less competitive at full leaderboard release than top-1 release, because it requires splitting the privacy budget across all of the $10$ fraud detectors benchmarked. In contrast, synthetic data methods only use up privacy budget once to generate the synthetic data and then can benchmark any number of fraud detectors with no further privacy loss.

\subsection{Subsample-and-Aggregate}
\label{app:subsample_aggregate}

\begin{figure}[ht]
    \centering
    \includegraphics[width=0.6\linewidth]{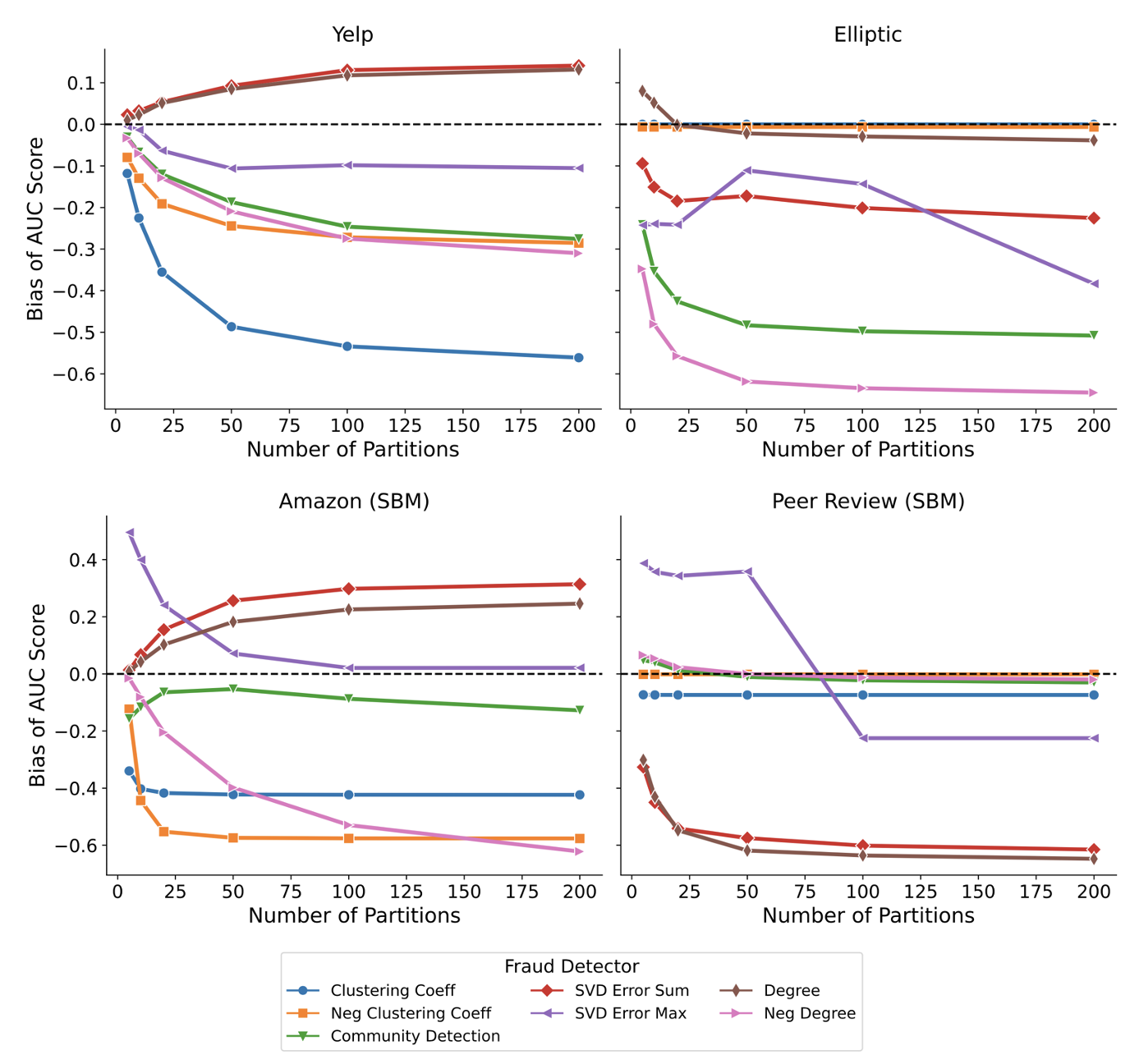}
    \caption{Bias to the AUC score introduced by subsample-and-aggregate for each fraud detector varying the number of partitions $(k)$ while fixing fraud sub-sampling rate of $\subsamplerate = 0.5$.}
    \label{fig:bias_vs_k}
\end{figure}

\begin{figure}[ht]
    \centering
    \includegraphics[width=0.6\linewidth]{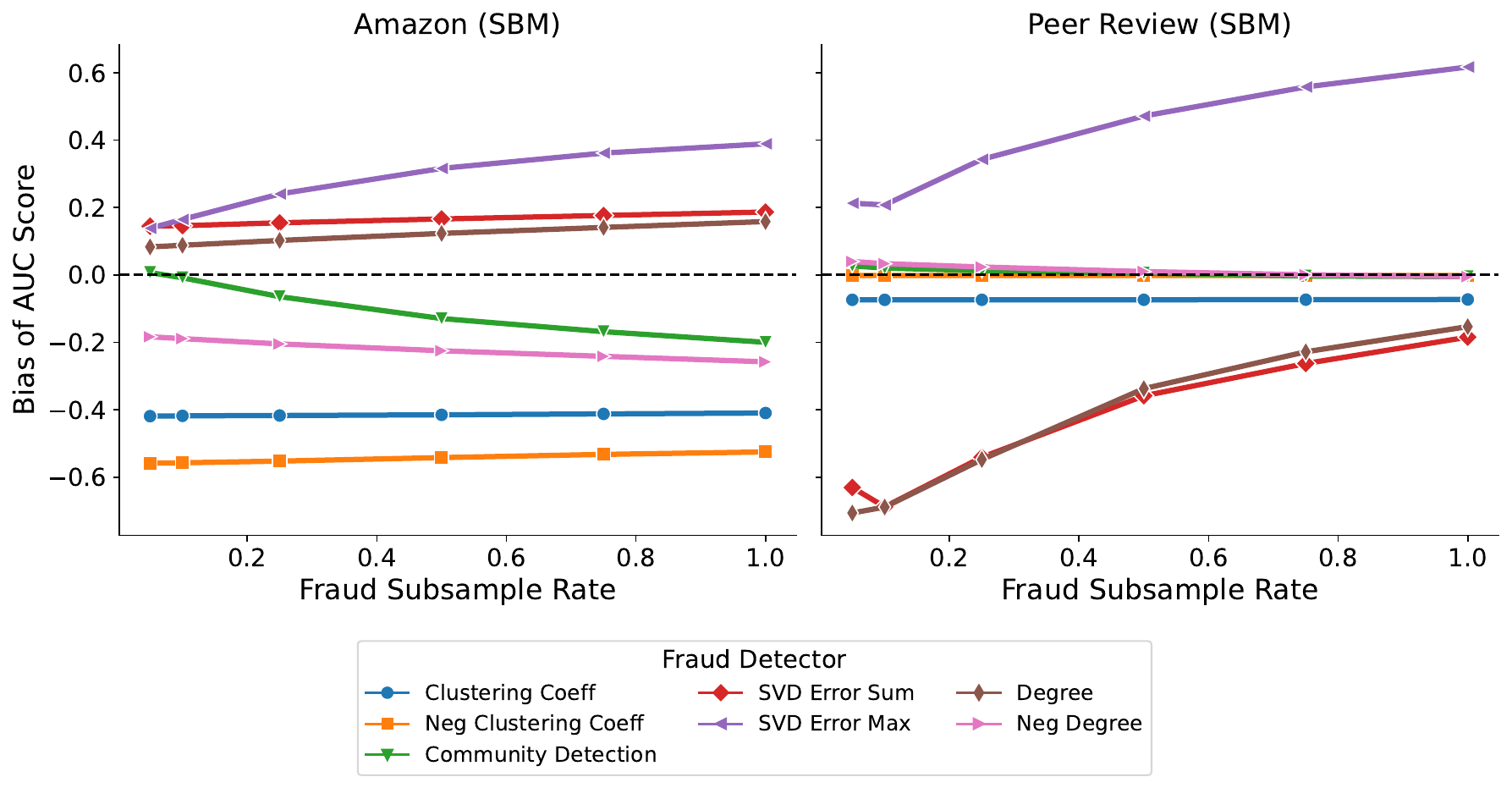}
    \caption{Bias to the AUC score introduced by subsample and aggregate for each fraud detector varying the fraud subsample rate $(\subsamplerate)$ while fixing number of partitions $\npartition = 20$.}
    \label{fig:bias_vs_subrate_otherdatasets}
\end{figure}

In this section, we give additional results for subsample and aggregate. First, in Figure~\ref{fig:bias_vs_subrate_otherdatasets}, we show the bias to different fraud detectors as a function of fraud subsampling rate for the Amazon and Peer Review datasets (as in Figure~\ref{fig:bias_vs_subrate} in the main text.) Then, in Figure~\ref{fig:bias_vs_k}, we show the bias of each fraud detector as a function of the number of partitions in subsample-and-aggregate. As expected, bias for each fraud detector increases in magnitude with more partitions (hence more graph distortion), but the sign and magnitude differ across different fraud detectors.

\subsection{Synthetic Data}
\label{app:synthetic_data}

In this section, we provide additional results for synthetic data methods. In Table~\ref{tab:stats_error}, we show the error to sufficient statistics using a more aggressive degree truncation threshold of $0.5$ times the max degree, compared to $1.0$ times the max degree in Figure~\ref{tab:suff_statistics} in the main text. Truncating more aggressively generally increases the error, except on Elliptic, where it decreases the error due to Elliptic being a highly sparse graph. 

\begin{table}
    \footnotesize
    \centering
    \begin{tabular}{l|c|c|c|c}
    \toprule
     & \# Edges & Degree Sequence & \# Triangles & Adjacency Matrix \\ 
    \midrule
    Yelp & 0.45 &  0.78 & 1.00 & 1.50 \\
    Elliptic & 17.61 & 25.64 & 25.13 & 1.91 \\
    Amazon (SBM) & 0.99 & 0.66 & 0.96 & 1.00 \\
    Peer Review (SBM) & 0.21  & 0.53 & 1.94 & 1.66 \\
    \bottomrule
    \end{tabular}
    \caption{Normalized mean absolute error (MAE) introduced to synthetic graph sufficient statistics at $\eps = 5.0$ and degree cutoff of $0.5$ the graph's max degree.}
    \label{tab:stats_error}
\end{table}

\begin{figure}[ht]
    \centering
    \includegraphics[width=0.7\linewidth]{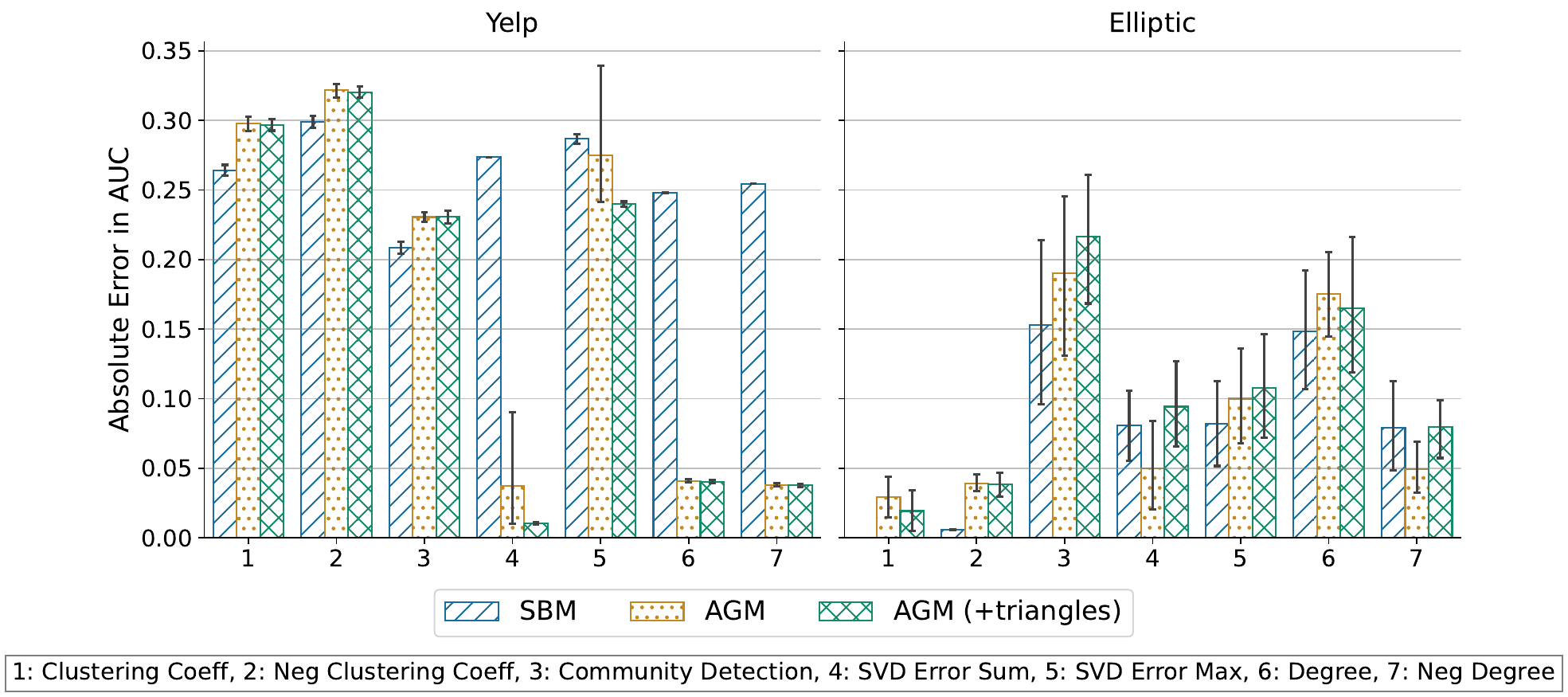}
    \caption{Comparison of average absolute error in AUC for each fraud detector for 10 synthetic graphs sampled based on sufficient statistics computed on the graph with no noise addition. All methods introduce significant bias in AUC estimates indicating that they do not capture important graph structure for fraud detection.}
    \label{fig:synthgraph_nonprivate_auc_err}
\end{figure}

\subsection{F1 Score}
\label{app:f1}

\begin{figure}[ht]
    \centering
    \begin{subfigure}[b]{0.48\linewidth}
        \centering
        \includegraphics[width=\linewidth]{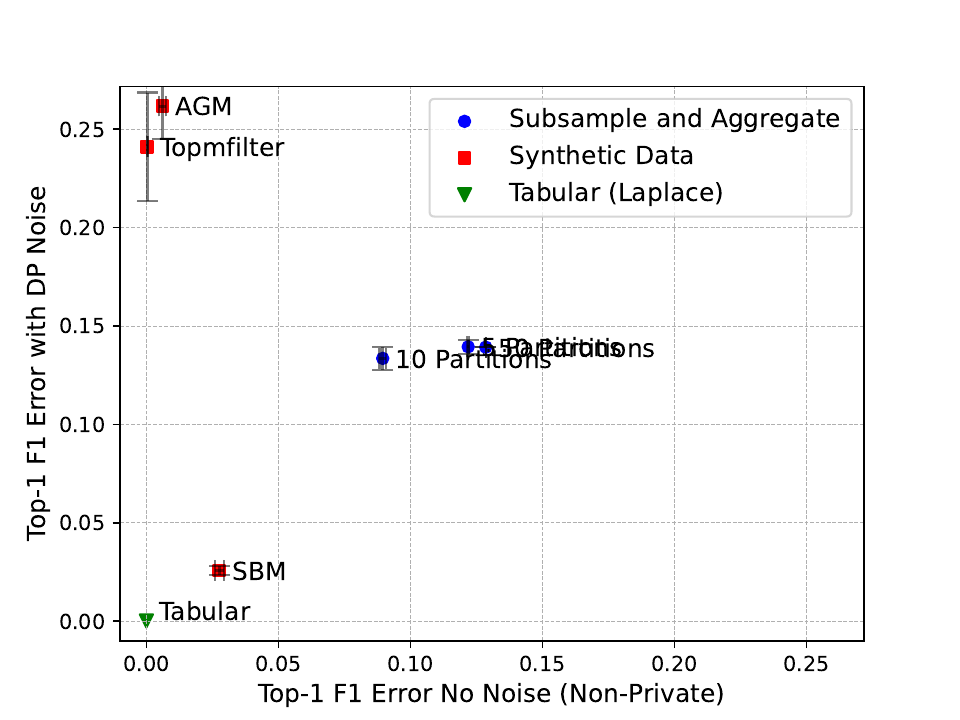}
        \caption{Amazon Data}
        \label{fig:amazon_f1}
    \end{subfigure}
    \hfill
    \begin{subfigure}[b]{0.48\linewidth}
        \centering
        \includegraphics[width=\linewidth]{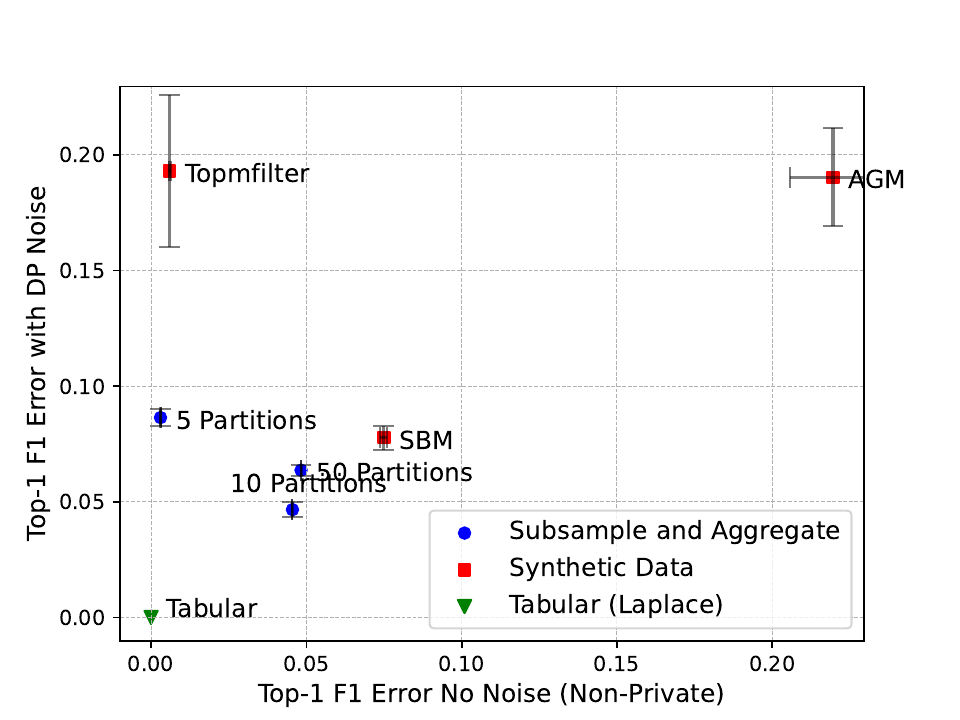}
        \caption{Yelp Data}
        \label{fig:yelp_f1}
    \end{subfigure}
    \caption{Top-1 F1 Score, $\eps=5.0$}
    \label{fig:amazon_yelp_f1}
\end{figure}

\begin{figure}[ht]
    \centering
    \begin{subfigure}[b]{0.48\linewidth}
        \centering
        \includegraphics[width=\linewidth]{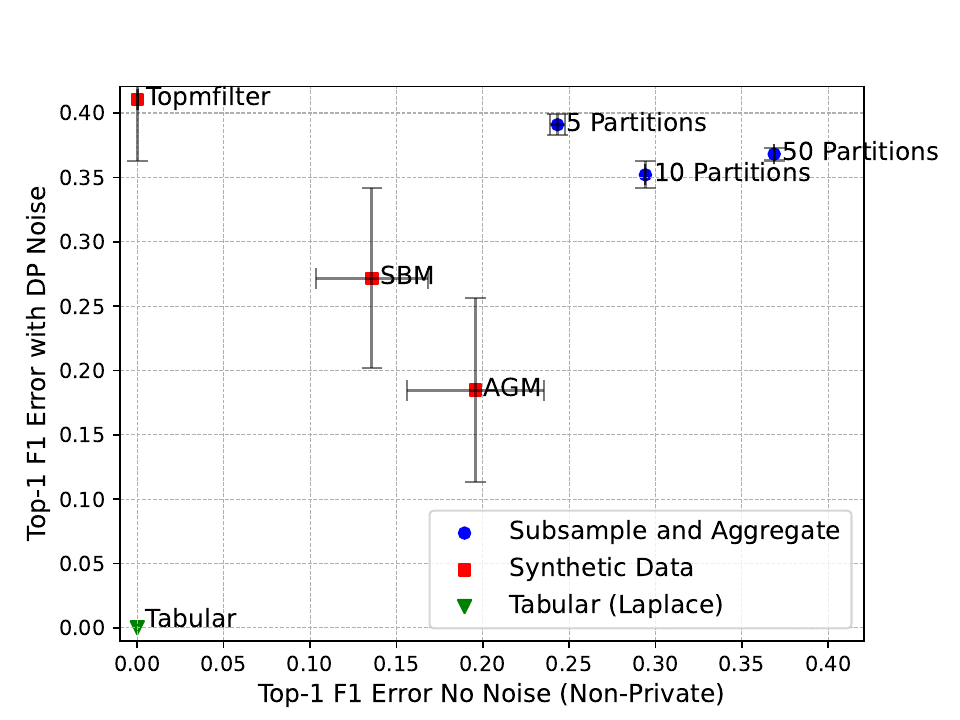}
        \caption{Elliptic Data}
        \label{fig:elliptic_f1}
    \end{subfigure}
    \hfill
    \begin{subfigure}[b]{0.48\linewidth}
        \centering
        \includegraphics[width=\linewidth]{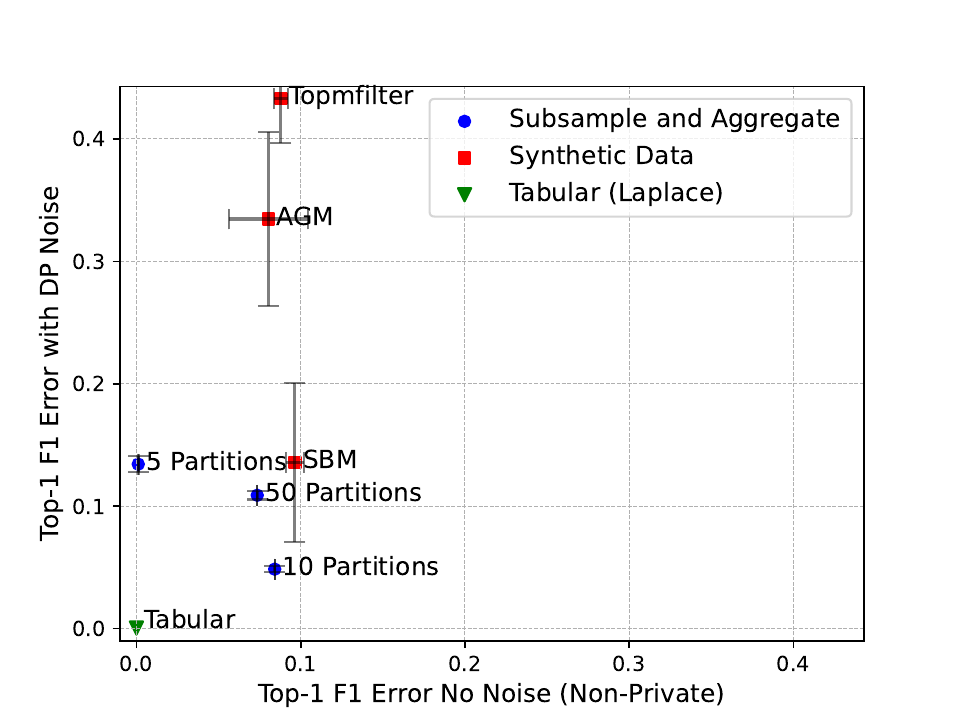}
        \caption{Peer Review Data}
        \label{fig:peer_review_f1}
    \end{subfigure}
    \caption{Top-1 F1 Score, $\eps=5.0$}
    \label{fig:elliptic_peer_review_f1}
\end{figure}

In this section, we give additional results using the F1 Score to benchmark fraud detectors instead of the AUC score. The F1 Score is the harmonic mean of the precision and recall of a classifier and has range $[0,1]$. As it depends on a threshold chosen to convert a fraud detection score into a fraud/benign label, for each fraud detector we compute the F1 score as the best F1 score across all possible thresholds. We show results for F1 score, analogous to Figure~\ref{fig:head_to_head_primary} in the main text, with $\eps=5.0$ for all datasets in Figures ~\ref{fig:amazon_yelp_f1} and \ref{fig:elliptic_peer_review_f1}.

\end{document}